\documentclass[a4paper,floatfix,amsmath,amssymb,prb,12pt]{revtex4-1}
\usepackage{amsmath}
\usepackage{graphicx}
\usepackage{epstopdf}
\usepackage{subfigure}
\usepackage[pdftex]{hyperref}
\usepackage{pstricks}

\usepackage{epsf}
\usepackage{dcolumn}
\usepackage{bm}
\usepackage{color}
\begin{document}
\newcommand{\newc}{\newcommand}
\newc{\mbf}{\mathbf}
\newc{\boma}{\boldmath}
\newc{\beq}{\begin{equation}}
\newc{\eeq}{\end{equation}}
\newc{\beqar}{\begin{eqnarray}}
\newc{\eeqar}{\end{eqnarray}}
\newc{\beqa}{\begin{eqnarray*}}
\newc{\eeqa}{\end{eqnarray*}}
\newc{\bd}{\begin{displaymath}}
\newc{\ed}{\end{displaymath}}

\title{Damped bead on a rotating circular hoop - a bifurcation zoo} 
\author{Shovan Dutta}
\affiliation{Department of Electronics and Telecommunication Engineering, Jadavpur University, Calcutta 700 032, India.}
\author{Subhankar Ray}
\email{sray.ju@gmail.com}
\affiliation{Department of Physics, Jadavpur University, Calcutta
700 032, India.}

\begin{abstract}
The evergreen problem of a bead on a rotating hoop shows a multitude of
bifurcations when the bead moves with friction.
This motion is studied for different values of the damping coefficient and rotational speeds
of the hoop. Phase portraits and trajectories corresponding to 
all different modes of motion of the bead are presented. They illustrate 
the rich dynamics associated with this simple system. 
For some range of values of the damping coefficient and rotational speeds
of the hoop, linear stability analysis of the equilibrium points
is inadequate to classify their nature. A technique involving transformation
of coordinates and order of magnitude arguments is presented to examine such
cases. This may provide a general framework to investigate other complex systems.
\end{abstract}
\maketitle

\section{Introduction}
\label{intro}
The motion of a bead 
on a rotating circular hoop\cite{goldstein}
shows several classes of fixed points and 
bifurcations\cite{jordansmith, strogatz,marsdenratiu}. It also exhibits 
reversibility, symmetry breaking, 
critical slowing down, homoclinic and heteroclinic orbits 
and trapping regions. It has been shown to provide a mechanical 
analogue of phase transitions\cite{fletcher}. It can also operate as a 
one-dimensional ponderomotive particle trap\cite{johnrab}. The rigid 
pendulum, with many applications, can be considered a special case of 
this system\cite{butikov1, butikov2}.

In this article we examine the motion of a damped bead on a rotating 
circular hoop. Damping alters the nature of the fixed points of the 
system, showing rich nonlinear features.
The overdamped case of this model\cite{strogatz,mancuso} and a 
variant involving dry friction\cite{burov} has been previously 
studied. 

For certain values of the damping coefficient 
and the rotational speed of the hoop, linear stability 
analysis predicts a line of fixed points and some of the fixed points 
appear as degenerate nodes.
However, such fixed points are borderline cases, sensitive to nonlinear
terms. 
By transforming to polar coordinates and employing order of magnitude 
arguments we analyze these borderline cases to determine the exact 
nature of these fixed points. 
To our knowledge, such analytical treatment does not appear in literature. 
The basic equations obtained for this system are quite generic and arise 
in other systems (e.g. electrical systems) as well. Hence,  
our technique may serve as a framework for investigating 
other more complex nonlinear systems.

\section{The physical system}
\label{phys}
\begin{figure}[h]
\centering
{\includegraphics[width=6cm]{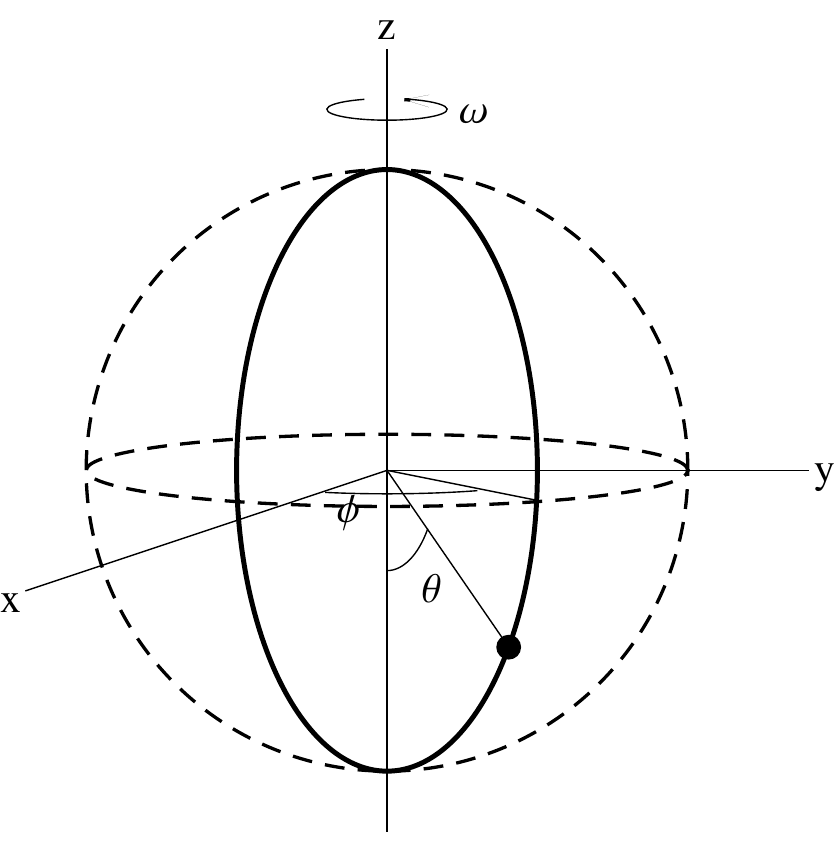}}
\caption{Schematic figure of a bead sliding on a rotating hoop showing the angles $\theta$ and $\phi$.} 
\label{bead_hoop}
\end{figure}

A bead of mass $m$, moves on a circular hoop of 
radius $a$. The hoop rotates
about its vertical diameter with a constant angular velocity $\omega$. 
The position of the bead on the hoop is given by angle $\theta$, 
measured from the vertically downward direction ($-z$ axis),
and $\phi$ is the angular displacement of the hoop from its initial
position on the $x$-axis (Figure \ref{bead_hoop}).

The Lagrangian of the system with no damping is,
\beq \label{lag1} 
L(\theta, \dot{\theta}) = \frac{ma^2}{2} (\dot{\theta}^2 +\omega^2 \sin^2 \theta )+ m g a \cos \theta 
\;, \; \; \mbox{where} \; \omega = \dot{\phi} \; \mbox{is a constant}.
\eeq
Using the Euler-Lagrange equation, the equation of motion is obtained as,
\beq 
\label{eom1}
\ddot{\theta} = \sin{\theta} (\omega^2 \cos \theta -g/a) .
\eeq
To include friction, a term $-b \dot{\theta}$ is introduced in (\ref{eom1})
as,
\beq \label{lag2}
\ddot{\theta} = \sin{\theta} (\omega^2 \cos \theta -g/a) -b\dot{\theta} \; ,
\eeq
where $b$ is the damping coefficient.
We identify $\omega^2_c= g/a$ as the critical speed of rotation of the hoop, 
and write $k = \omega^2 /\omega^2_c$, $\mu = b/\omega_c$.
Defining $\tau = \omega_c \; t$, (\ref{lag2}) may be made 
dimensionless by changing from $t$ to $\tau$, 
\beq \label{eom2}
{\theta}^{''} = \sin{\theta} (k \cos \theta -1) -\mu \theta^{'}, 
\;\; \mbox{where} \; \frac{d ^2 \theta}{d \tau ^2}={\theta}^{''} 
\; .
\eeq
For phase plane analysis, we define a new variable $\theta_1 = \theta^{'}$,
and write (\ref{eom2}) as,
\beqar
\theta^{'} &=& \theta_1 \hspace{4.4cm} \label{fo1} \\ 
\theta_1{'}  &=& -\sin\theta (1 - k \cos\theta ) - \mu \;\theta_1 \; .
\label{fo2}
\eeqar
The parameter $k$ can take only positive values whereas $\mu$ may be either
positive or negative.

Due to the symmetry of the hoop about its vertical axis, (\ref{fo1})
and (\ref{fo2}) remain invariant under the transformations 
$\theta \to - \theta \; , \; \theta_1 \to - \theta_1$. This
implies that alternate quadrants of the $\theta-\theta_1$ plane have 
similar trajectories. Similarly, it is easily verified that if
($\theta(t), \theta_1(t)$) is a solution for positive damping 
($\mu > 0$), then for negative damping ($\mu < 0$), 
($\theta(-t) , -\theta_1(-t)$) and ($-\theta(-t) , \theta_1(-t)$) 
are two solutions. The phase portrait of the system for negative damping will just be 
the reflection of the positive damping phase portrait with 
the arrows reversed. Hence we confine our attention to $\theta \in [0,\pi]$
and $\mu > 0$.

When there is no damping\cite{shovan1}, the fixed points are at $(0,0)$ and 
$(\pi,0)$ for $0 \le k \le 1$, whereas for $k >1$, an additional 
fixed point appears at $(\Omega_1 = \cos^{-1}(1/k),0)$.
Damping changes the nature of fixed points and not their number or location.

\section{Nature of the fixed point $(0,0)$}
\label{fpoint1}
The Jacobian matrix at $(0,0)$ is obtained by Taylor expanding 
(\ref{fo1}) and (\ref{fo2}) about $(0,0)$ and retaining the linear terms.
\begin{equation}
\mathbf{J}(0,0) = \left(
\begin{array}{ccc}
0 & 1 &  \\
k-1 & -\mu 
\end{array} \right) \; .
\end{equation}

Let $\Gamma$ and $\Delta$ denote the trace and determinant of
the above matrix.

\begin{figure}[h]
\centering
\subfigure[~stable spiral, $ \, \mu < 2 \sqrt{1-k}$]
{\includegraphics[width=2.8in]{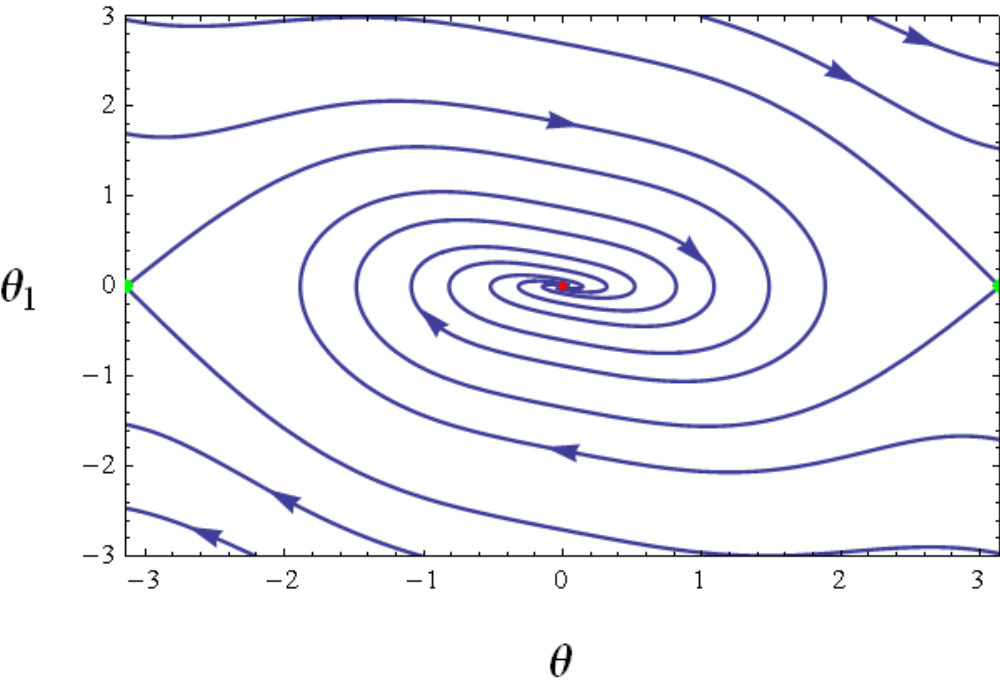}}
\hspace{1cm}
\subfigure[~ stable node, $\mu > 2 \sqrt{1-k}$]
{\includegraphics[width=2.8in]{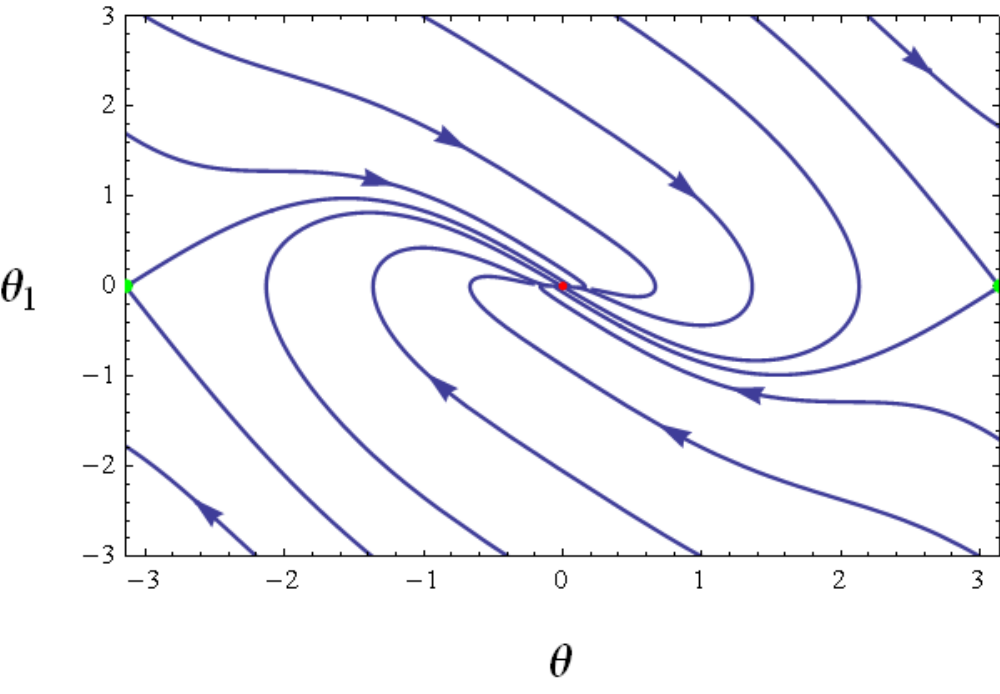}}
\caption{Phase trajectories around $(0,0)$ for $k =0.91$ showing (a) 
stable spiral for $\mu = 0.3$ and (b) stable node for $ \mu = 1$.}
\label{sspiral0}
\end{figure}

\begin{enumerate}
\item When $k>1$, both $\Gamma$ and $\Delta$ are negative.
The fixed point is a saddle with eigenvalues and eigenvectors given by,
\beq
\lambda_{1,2} = \frac{-\mu \pm \xi_1 }{2},\;\;\;
\mathbf{v}_{1,2} = \left(
\begin{array}{c}
1 \\
(-\mu \pm \xi_1 )/2
\end{array} \right) \; ,
\label{lam12}
\eeq
where $\xi_1 = \sqrt{\mu^2 + 4(k-1)}$. 
Saddles are robust and do not get perturbed by nonlinearities. Thus, 
$(0,0)$ will remain a saddle even if nonlinear terms are taken into account 
(see Figures \ref{spiralomega1}, \ref{nodeomega1} and \ref{Hopf2}).

For $\mu=0$, both $\lambda_1$ and $\lambda_2$ equal $\sqrt{k-1}$. 
As $k\to 1^+$, $\lambda_1 \to 0^+$ and $\lambda_2 \to -\mu^-$, which means 
that the saddle will start looking like a line of fixed points along the 
direction of $\mathbf{v}_1$ with solutions decaying along $\mathbf{v}_2$.

\item For $ 0 \le k < 1$, $\Gamma = -\mu$ 
is negative, whereas $\Delta =1-k$ is positive. When there is no damping, the
point $(0,0)$ is a center. As $\mu$ is increased, the center transforms into
a stable spiral for $\mu < 2\sqrt{1-k}$ as shown in Figure \ref{sspiral0}(a). 
The frequency of spiralling is $\nu \approx \sqrt{1-k-\mu^2/4}$. 
As $\mu\to2\sqrt{1-k}^-$, $\nu\to0^+$. For $\mu > 2\sqrt{1-k}$, 
the fixed point $(0,0)$ transforms to 
a stable node (Figure \ref{sspiral0}(b)). For $\mu = 2\sqrt{1-k}$, it is a
degenerate node. However, degenerate nodes are 
borderline cases and are sensitive to nonlinear terms.

\item For $k=1$, (\ref{fo1}) and (\ref{fo2}) simplify to,
\beqar
\theta^{'} = \theta_1 \hspace{4.2cm} 
\label{k1fo1}\\ 
\theta_1^{'}  = -\sin\theta (1 - \cos\theta ) - \mu\;\theta_1 \; .
\label{k1fo2}
\eeqar
In the linearized dynamics, $\theta_1$ decays exponentially 
as $e^{-\mu \,t}$. In the phase plane, all trajectories move 
along a straight line with slope $-\mu$ and 
stop on reaching the $\theta$ axis. 
However, the inclusion of nonlinear terms changes this situation.
\end{enumerate}
\subsection{Nature of $(0,0)$ with nonlinearities}
\label{nlin0}

{\bf 1.} $\mathbf{0 \le k < 1}$ and $\mathbf{\mu < 2\sqrt{1-k}}$ :

To include the effect of nonlinear terms, let us define two new variables,
\beq
\label{rphi}
\theta = r \cos\phi,\;\;\;\theta_1 = r \sin\phi
\eeq
Equations (\ref{fo1}) and (\ref{fo2}) then may be written as,
\beqar
r^{'} = r[\cos \phi \sin \phi - 2\sqrt{1-k}\sin^2 \phi] - \sin \phi \sin (r\cos \phi)[1 - k\cos (r\cos \phi)] 
\label{req} \\
\phi^{'} = -\sqrt{1-k}\sin (2\phi) - \sin^2 \phi - \frac{\cos \phi}{r} \sin (r\cos \phi)[1 - k\cos (r\cos \phi)] \hspace{0.8cm}
\label{phieq}
\eeqar
We wish to examine the fixed point(s) in the $r-\phi$ plane
corresponding to $(0,0)$ in the $\theta-\theta_1$ plane, to determine their
true nature.
Strictly speaking, (\ref{req}) and (\ref{phieq}) are meaningful
only when $r > 0$. 
Neither $\phi$ nor $\phi^{'}$ have any meaning when $r=0$.
Hence, we may assign any arbitrary function $f(\phi)$ to $\phi^{'}$ at $r=0$ 
without altering physical predictions. 
However, (\ref{phieq}) describes accurately 
the approach to $r=0$ (if any) in the $r-\phi$ plane at 
arbitrarily small scales. We therefore set $f(\phi)$ equal to 
the limiting value of $\phi^{'}$ as $r\to0$.
\beq
f(\phi) = \lim_{r \rightarrow 0}\phi^{'} = -\sqrt{1-k}\sin (2\phi) 
+ \frac{k}{2}\cos (2\phi) - \big(1-\frac{k}{2}\big)
\eeq
Equations (\ref{req}) and (\ref{phieq})
are periodic in $\phi$ with period $\pi$. Hence, 
the phase portrait in the $r-\phi$ plane is periodic along the $\phi$-axis
with period $\pi$. This means that in the $\theta-\theta_1$ plane,
the phase space is symmetric about $(0,0)$.

Using the identity $(\sqrt{1-k})^2 + (k/2)^2 = (1-k/2)^2$, one can write 
$f(\phi)$ in the form,
\beq
f(\phi) = \left(1-\frac{k}{2}\right)
[\cos (2 \phi + \alpha)-1] \;,
\label{1phistar}
\eeq
where $\alpha = 2 \tan^{-1}\sqrt{1 - k}$, $n=0,1,2, \dots$. Therefore, 
fixed points $(0,\phi^{*})$ in the $r-\phi$ plane, where $f(\phi^{*}) = 0$,
are given by $\phi^{*}_n = n\pi - \tan^{-1}\sqrt{1-k}$ with $n=0,1,2, \dots$.
These correspond to the point $(0,0)$ in 
the $\theta-\theta_1$ plane.

The fixed points $(0, \phi^{*}_n)$ in the $r-\phi$ plane are
separated by $n \pi$ (where $n$ is any integer). Hence, in
the $\theta-\theta_1$ plane, there are no trajectories that 
can approach $(0,0)$ along two independent directions.
So we can say that $(0,0)$ in the $\theta-\theta_1$ plane
cannot be a stable node. In the $r-\phi$ plane, close to some fixed point $(0,\phi_n^*)$, if
there exist trajectories that approach this point and stop there,
the corresponding fixed point $(0,0)$ in the $\theta-\theta_1$ plane
cannot be a stable spiral. For a spiral, $\phi \to \pm \infty$
as $r \to 0$.
As (\ref{req}) and (\ref{phieq}) are periodic in $\phi$ with period $\pi$, 
the nature of all fixed points on the $\phi$-axis separated by $\pi$ is
identical. So we may choose to investigate 
$(0,\phi^{*}=-\tan^{-1}\sqrt{1-k})$. Linearization about this point incorrectly predicts the whole $\phi$ axis to be a line of fixed points. So, we must include the effects of the nonlinear terms. Let $\epsilon = \phi-\phi^{*}$. For small $r$, we may write (\ref{req}) and (\ref{phieq}) as,
\beqar
r^{'} &=& -r\Big[\sqrt{1-k} - \big(1-\frac{k}{2}\big)\sin 
(2\epsilon)\Big] + O(r^3) 
\label{reps} \\
\epsilon^{'} &=& -\left(1-\frac{k}{2}\right)[1-\cos (2\epsilon)]
+ O(r^2) \hspace{1.6cm}
\label{eps}
\eeqar
Consider an initial condition, $r(\tau=0)=r_0$ and $\epsilon(\tau=0)
=\epsilon_0$, where $0<\epsilon_0<\tan^{-1}(\sqrt{1-k}/2)$. 
For any finite positive value of $1-k$, $\epsilon_0$ is finite and $<1/2$. 
Therefore, $r_0$ may be chosen sufficiently small, $0<r_0\ll\epsilon_0$, 
so as to make all terms of $O(r^2)$ and higher, negligible compared to the 
leading terms in (\ref{reps}) and (\ref{eps}). When these terms are 
neglected, (\ref{reps}) and (\ref{eps}) may be solved to yield,
\beqar
\epsilon(\tau) &=& \tan^{-1} \Big[\frac{1}{(2-k)\tau + \cot 
\epsilon_0}\Big] \hspace{4.2cm} \label{etau} \\
r(\tau) &=& r_0 \sin \epsilon_0 \sqrt{[(2-k)\tau + \cot \epsilon_0]^2 +1} \hspace{0.1cm} \exp(-\sqrt{1-k}\;\tau)
\label{rtau}
\eeqar
According to this solution, the trajectory monotonically 
approaches the point $(0,0)$ in the $r-\epsilon$ plane as $\tau\to\infty$. 
This behaviour will hold even with inclusion of nonlinear terms, provided,
the terms independent of $r$ and of $O(r)$ remain dominant
over the entire trajectory. The following arguments establish that it is
indeed so.

First, the trajectory cannot reach the $\phi$ axis at a positive value of 
$\epsilon$. This is because on the $\phi$ axis, $r'=0$ and 
$\phi^{'} = f(\phi) < 0$ in between two fixed points. Thus, there is 
already a trajectory running along the $\phi$ axis directed 
towards $\epsilon =0$.

For the initial point $(r_0,\epsilon_0)$, with the choice 
$0<r_0\ll\epsilon_0<1/2$, both $r$ and $\epsilon$ will start to decrease
as per (\ref{reps}) and (\ref{eps}). Hence $r^2$ will become 
more negligible compared to $r$. Also, for all $0\leq\epsilon\leq\epsilon_0$,
the $\epsilon$ independent term, namely, $-\sqrt{1-k} \, r$, 
in (\ref{reps}), will be dominant.
However, if $\epsilon$ decays more rapidly, such that at some stage 
$r \sim\epsilon$, then the $O(r^2)$ term will contribute on 
the same scale as the first term in (\ref{eps}), which is $O(\epsilon^2)$. Similarly, the $O(r^3)$ term will contribute on the same scale as the 2nd term in (\ref{reps}) if in the course of decay, at some point $\epsilon \sim r^2$. 
However, such situations will never arise as is shown below.

Let us assume that $r_0 = \epsilon_0^{20} \ll 1$ and that $\epsilon$ decreases
very rapidly, such that, at some instance, $r=\epsilon^2$. 
As $r$ has decreased monotonically from its initial value, we must have 
$\epsilon=r^{1/2}<r_0^{1/2}$. However, $r_0^{1/2}=\epsilon_0^{10}\ll1$. Hence, 
along the entire trajectory, up to this instance, terms of 
$O(r^2)$ and higher are negligible compared to the leading terms in 
(\ref{reps}) and (\ref{eps}).
Thus the solutions (\ref{etau}) and (\ref{rtau}) are valid and give the 
correct orders of magnitudes of the dynamical quantities.

As $0 < \epsilon < \epsilon_0^{10} \ll 1$ and $0 < \epsilon_0 < 1/2$, we may write, $\tan \epsilon 
\sim\epsilon$ and $\tan \epsilon_0\sim\epsilon_0$. 
Hence, $0<\tan \epsilon<\tan^{10} \epsilon_0$, which implies
$\tan \epsilon <(\sqrt{1-k}/2)^{10}\ll1$. Combining this with (\ref{etau}) 
and using the fact that $(2-k)\sim 1$, we get $\tau \gtrsim 
(\cot\epsilon_0)^{10}$, a very large quantity. 
Meanwhile (\ref{etau}) and (\ref{rtau}) together imply,
\begin{equation}
\frac{r}{\epsilon^2} \sim r_0 \sin\epsilon_0 \Big[\frac{\tau^3}{\exp(\sqrt{1 - k}\;\tau)} \Big] 
\end{equation}
Both $r_0\sin\epsilon_0$ and the quantity within brackets are $\ll1$, 
which implies that $r\ll\epsilon^2$. This is in contradiction to the initial 
assumption that $r=\epsilon^2$. Therefore, we conclude that starting with 
the prescribed initial condition, $r$ will never become equal to $\epsilon^2$. 
This ensures that along the entire trajectory, 
the terms independent of $r$ and of $O(r)$ remain the dominant terms
in (\ref{reps}) and (\ref{eps}).
Both $r$ and $\epsilon$ will decrease 
monotonously toward their respective zero values. Neither can the trajectory cross the curve $r=\epsilon^2$ nor reach the $\phi$ axis before 
$\epsilon$ becomes zero. Thus, the trajectories in the $r-\epsilon$ plane
must approach $(0,0)$ tangential to the $\phi$ axis and slow 
to a halt there.

\begin{figure}[h]
\centering
\subfigure[~stable degenerate node, $\mu = 0.6$]
{\includegraphics[width=2.8in]{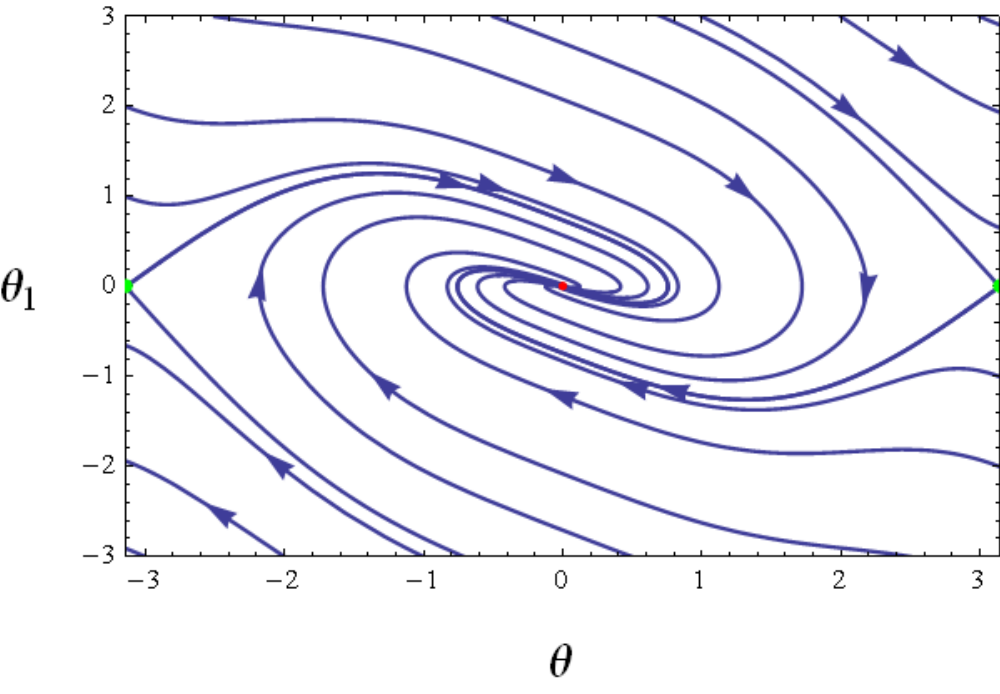}}
\hspace{1cm}
\subfigure[~central region magnified]
{\includegraphics[width=3in]{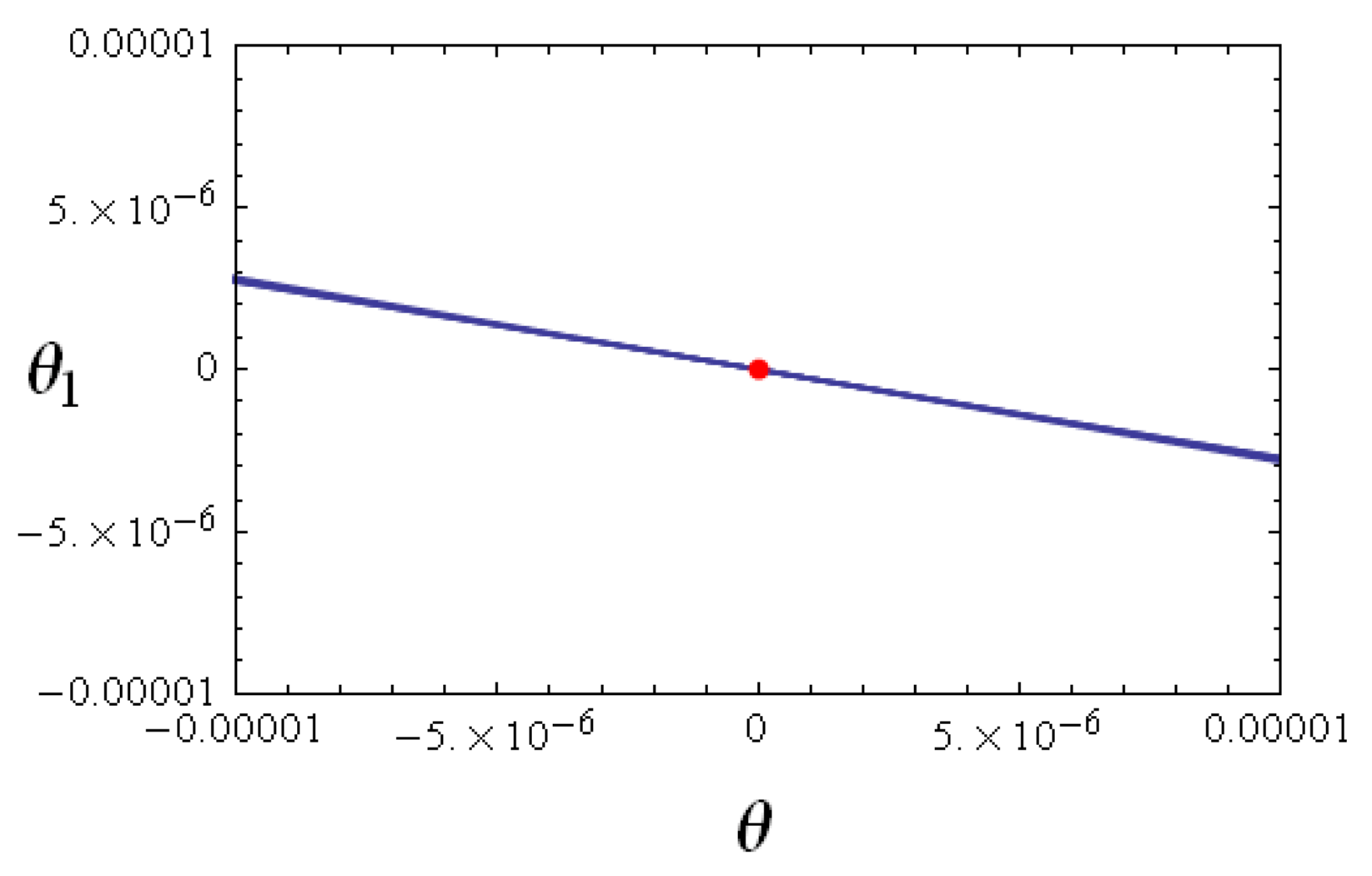}}
\caption{Phase trajectories around $(0,0)$ for $k=0.91$ and 
$\mu = 2\sqrt{1 - k}$.}
\label{deg}
\end{figure}

In the $\theta-\theta_1$ plane, the above arguments imply that, 
trajectories exist which start at a 
finite distance from $(0,0)$ and reach this point along a line of slope 
$-\sqrt{1 - k}$. Also, no other such line with a different slope exists. 
These facts clearly establish that $(0,0)$ is a stable degenerate node 
(Figure \ref{deg}). 

{\bf 2.} $\mathbf{k = 1}$ :

Proceeding as before, (\ref{k1fo1}) and (\ref{k1fo2}) may be 
written in terms of $r$ and $\phi$ as,
\beqar
r^{'} &=& r[\cos \phi \sin \phi - \mu \sin^2 \phi] - \sin \phi \sin 
(r\cos \phi)[1 - \cos (r\cos \phi)] 
\label{k1req} \\
\phi^{'} &=& -\mu \cos \phi \sin \phi - \sin^2 \phi - \frac{\cos \phi}{r} 
\sin (r\cos \phi)[1 - \cos (r\cos \phi)]
\label{k1phieq}
\eeqar

$f(\phi)$ is defined as,
\beq
f(\phi) = \lim_{r \rightarrow 0}\phi^{'} 
= -\mu \cos \phi \sin \phi - \sin^2 \phi
\eeq

The phase portrait is periodic in $\phi$ with period $\pi$. 
The fixed points in the $r-\phi$ plane of the form (0,\, $\phi^*$)
are given by,
\beq
\phi^*_1 = n\pi \hspace{0.5cm}\mbox{and}\hspace{0.5cm} \phi^*_2 = n\pi 
- \tan^{-1}\mu \;\; n=0,1,2,\dots  .
\eeq

For $-\tan^{-1}\mu<\phi<0$, $f(\phi)>0$ whereas for 
$0<\phi<\pi - \tan^{-1}\mu$, $f(\phi)<0$. The positive and negative
nature of $f(\phi)$ repeats periodically along the $\phi$ axis. 

The Jacobian at the points 
$(0,n\pi - \tan^{-1}\mu)$ given by,
\begin{equation}
\mathbf{J}(0,\phi_2) = \left(
\begin{array}{ccc}
-\mu & 0 &  \\
0 & \mu 
\end{array} \right)
\end{equation}
is traceless and has a negative determinant $\Delta=-\mu^2$. So, this 
family of fixed points are saddles having stable manifold along $r$ axis 
and unstable manifold along $\phi$ axis.

Linear analysis of the family of fixed points $(0,n\pi)$, 
incorrectly predicts $\phi=n\pi$ to be lines of 
fixed points. 
Let us examine the point $(0,0)$ in the $r-\phi$ plane 
for simplicity. Consider the condition,
\beq
\label{cond}
0\leq |\phi| \leq r \ll min\{1,\mu\}
\eeq
If (\ref{cond}) holds, then neglecting terms of $O(r^3)$ and smaller
in (\ref{k1req}) and (\ref{k1phieq}), we may write,
\beqar
r^{'} &=& \phi r + \eta_1 \hspace{2cm} \label{k1r2} \\
\phi^{'} &=& -\mu \phi - \frac{r^2}{2} - \phi^2 + \eta_2
\label{k1phi2}
\eeqar
where $\eta_1 \sim \phi^2 r$ and $\eta_2 \sim \phi^3 \;\mbox{or}\; 
\phi^2 r^2$, whichever is larger. Note that as long as (\ref{cond}) 
is satisfied, both $\eta_1$ and $\eta_2$ can at most be of the 
order of $r^3$.

Let us take the initial condition $\phi_0 = 0$ and $0<r_0\ll 1$. 
Then, $\phi^{'} (t=0) = -r_0^2/2$ and $r^{'} (t=0) = 0$. 
Hence, $\phi$ will start decreasing and become negative. As a result, 
$r^{'}$ will become negative and remain so until $\phi$ or $r$ vanishes, 
provided (\ref{cond}) remains true. It is seen that as long 
as the trajectory is above the curve $\phi=-r^2/2\mu$, (\ref{cond}) is 
satisfied and both $r^{'}$ and $\phi^{'}$ are negative. Therefore, 
the trajectory approaches and eventually crosses this curve, where 
$\phi^{'}$ is still negative, being of the order of $\phi^2$. 

\begin{figure}[h]
\centering
\includegraphics[height=6cm]{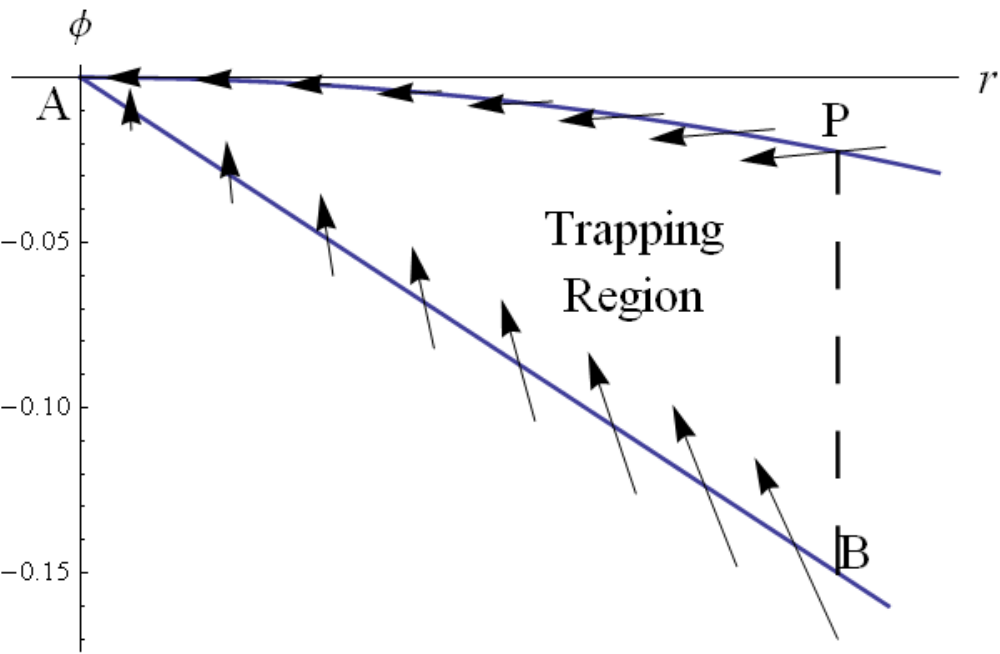}
\caption{Trapping Region.}
\label{trap}
\end{figure}

Let us consider the `trapping region' in Figure \ref{trap}, which shows the 
phase flow on the curves $\phi=-r$ and $\phi=-r^2/2\mu$.
At any point on the line $\phi=-r$ for which $r \ll min\{1,\mu\}$, 
\begin{equation}
\nonumber \frac{\phi^{'}}{r^{'}} = -\frac{\mu}{r} + O(1)
\end{equation}
As $r\ll \mu$, $|\phi^{'}/r^{'}| \gg 1$. Thus the phase flow is 
almost vertically upward, as shown in Figure \ref{trap}. Everywhere inside the 
region, (\ref{cond}) is satisfied and hence $r^{'} < 0$ and finite. 
Consequently, after entering the region at point $P$, the trajectory 
must constantly move towards left. Again, it cannot penetrate the curves 
$AP$ or $AB$, because other trajectories are actually flowing inward 
across them. Hence, we have trapped it. Upon arrival at any point on the 
arc $AP$, a trajectory must inevitably land up at $(0,0)$. 
Note that at any point on the line $\phi=-mr$, $(0 < m \leq 1)$,
\begin{equation}
\nonumber \frac{\phi^{'}}{r^{'}} = -\frac{m\mu - (m^2 + 1/2) r + 
O(m^3 r^2)}{mr + O(m^2 r^2)} \; .
\end{equation}
For any finite value of $m$, this approaches $-\infty$ as $r\to 0$, meaning 
that the phase flow is almost vertically upward on any line of non-zero 
slope near $(0,0)$. Therefore, the trajectory must reach $(0,0)$ along 
the $r$ axis.

Thus, the fixed points $(0,n\pi)$ in the $r-\phi$ plane are stable nodes 
having slow eigenvector along $r$ axis and fast eigenvector along $\phi$ 
axis. In between these, lie the saddle points $(0,n\pi - \tan^{-1} \mu$) (Figure \ref{rphi}).
\begin{figure}[h]
\centering
\subfigure[~stable node and saddle points]
{\includegraphics[width=2.8in]{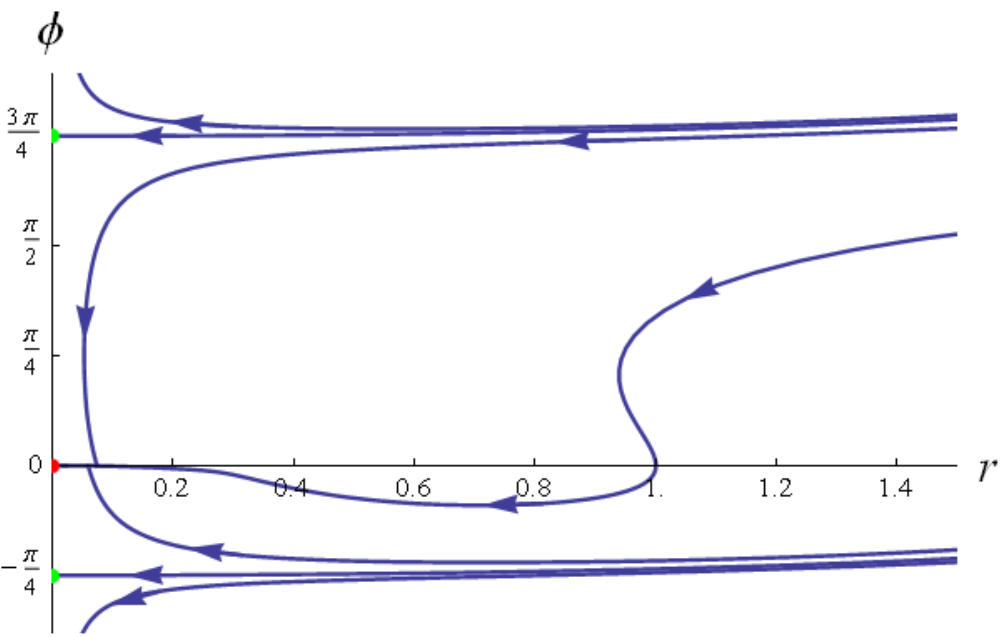}}
\hspace{1cm}
\subfigure[~region near $(0,0)$ magnified]
{\includegraphics[width=2.8in]{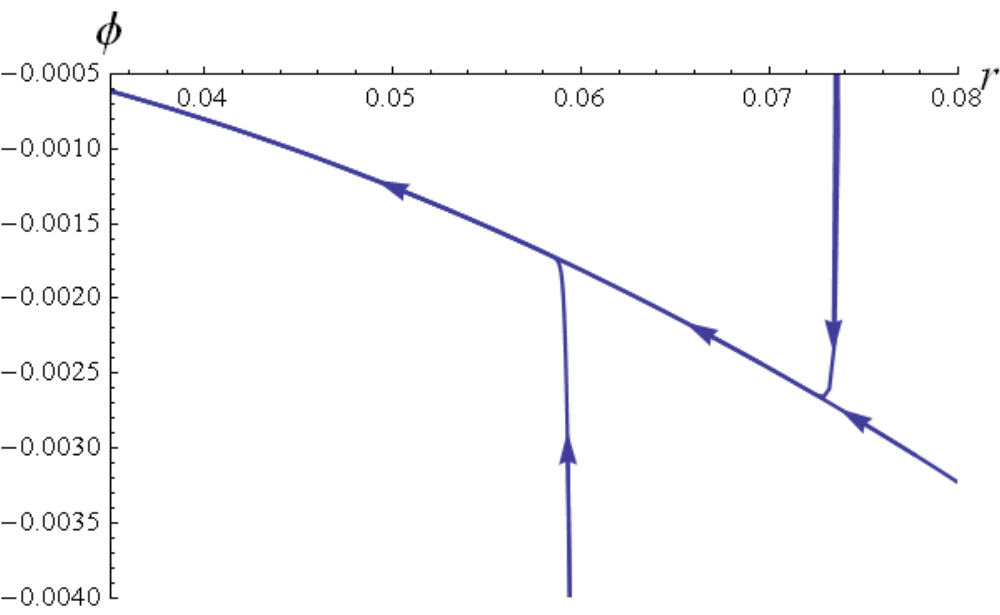}}
\caption{Phase portrait in the $r-\phi$ plane.}
\label{rphi}
\end{figure}

These results from the $r-\phi$ plane mean that in the $\theta-\theta_1$ 
plane, two trajectories exist which reach $(0,0)$ along the line of slope 
$-\mu$ and all other neighbouring trajectories reach it along the 
$\theta$ axis. In other words, $(0,0)$ is a stable node here.
\begin{figure}[h]
\centering
\subfigure[~stable node]
{\includegraphics[width=2.8in]{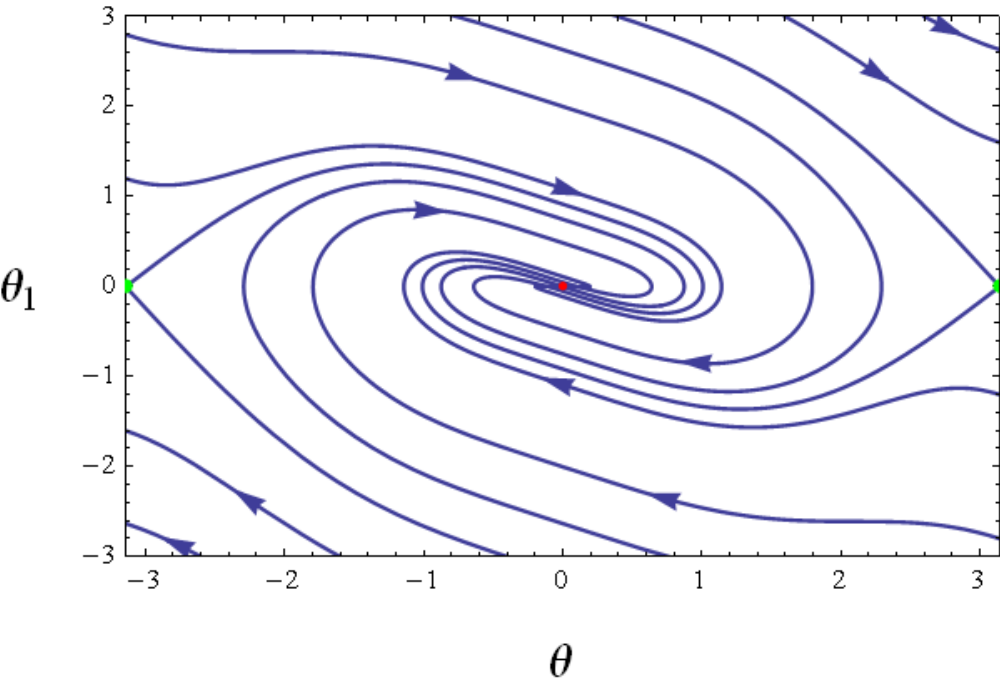}}
\hspace{1cm}
\subfigure[~central region magnified]
{\includegraphics[width=2.8in]{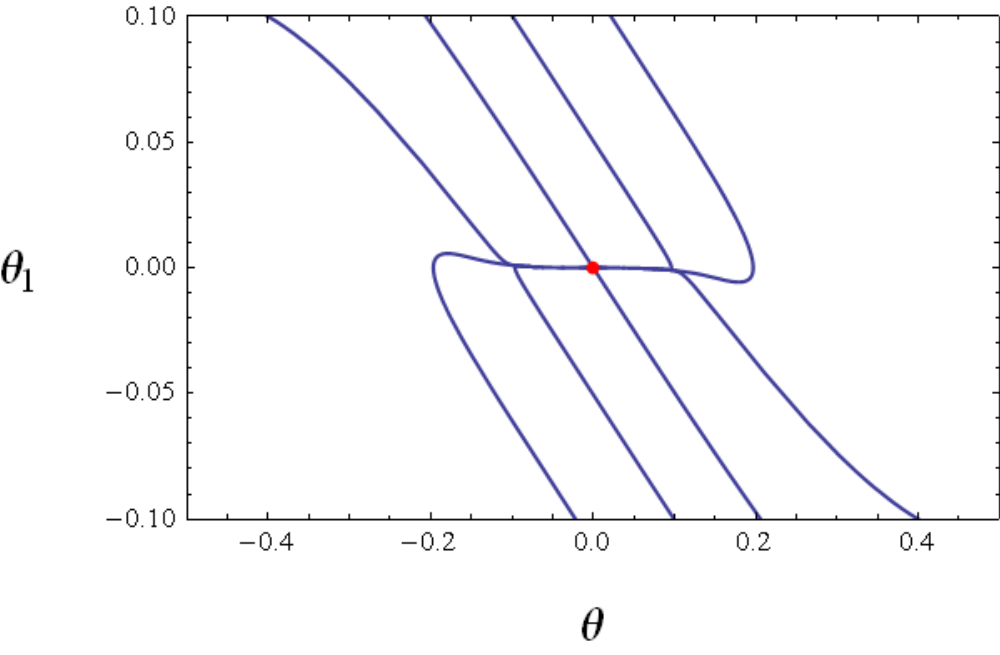}}
\caption{Phase trajectories about $(0,0)$ in the $\theta-\theta_1$
plane for $k=1$ and $\mu=0.5$.}
\label{stablenode}
\end{figure}

\begin{figure}[!h]
\centering
\subfigure[~$(0,0)$, curve is $\mu=2\sqrt{1 - k}$]
{\includegraphics[width=2.8in]{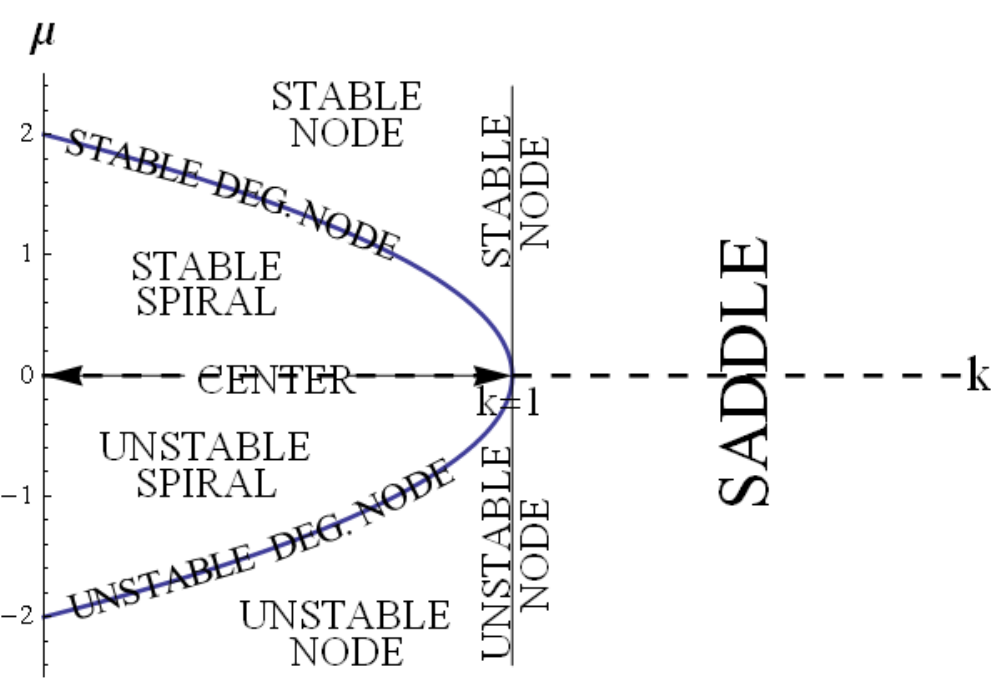}}
\hspace{1cm}
\subfigure[~$(\pm \Omega_1,0)$, curve is $\mu=2\sqrt{k-1/k}$]
{\includegraphics[width=2.8in]{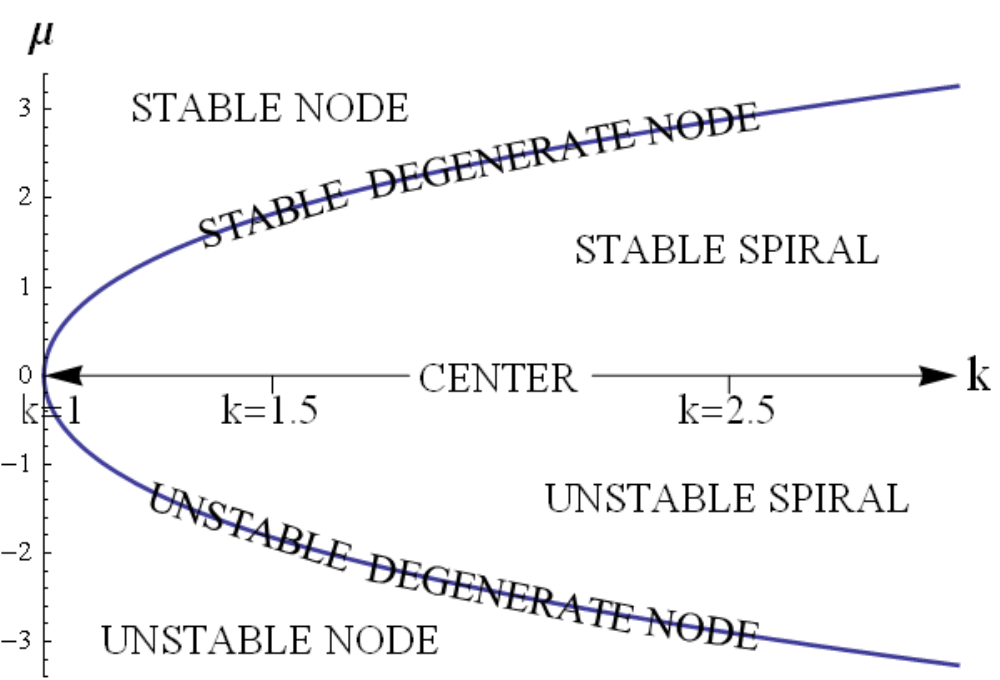}}
\caption{Nature of fixed points $(0,0)$ and $(\pm \Omega_1,0)$ 
over the $k-\mu$ plane}
\label{param}
\end{figure}

Figure \ref{param}a shows the nature of the fixed point at $(0,0)$ over 
the entire parameter space.
\section{Nature of the fixed point $(\Omega_1,0)$}
\label{omega}
This fixed point exists when $k \ge 1$.
Linearization of (\ref{fo1}) and (\ref{fo2}) gives us the Jacobian
at $(\Omega_1,0)$ as,
\begin{equation}
\mathbf{J}(\Omega_1,0) = \left(
\begin{array}{ccc}
0 & 1 &  \\
\left(\frac{1}{k} - k\right) & -\mu 
\end{array} \right)
\end{equation}

\begin{figure}[h]
\centering
\subfigure[~center, $\mu=0$]
{\includegraphics[width=2.8in]{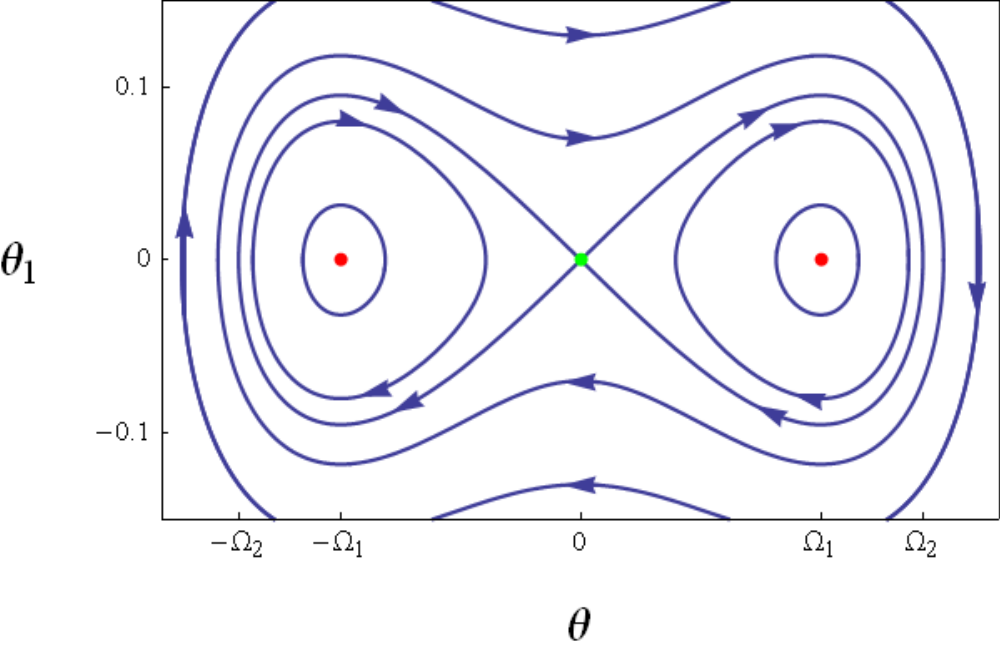}}
\hspace{1cm}
\subfigure[~stable spiral, $\mu=0.1$]
{\includegraphics[width=2.8in]{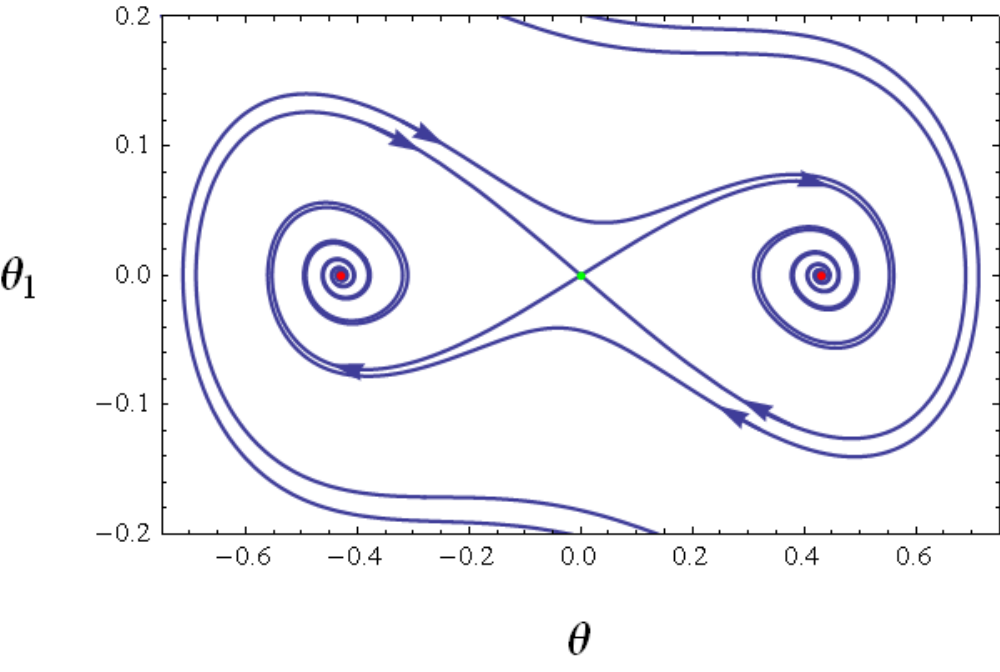}}
\caption{Phase trajectories around $(0,0)$ for $k=1.1$ and $0 \leq \mu < 2\sqrt{k - \frac{1}{k}}$.}
\label{spiralomega1}
\end{figure}

\begin{figure}[h]
\centering
\subfigure[~stable node, $\mu = 1$]
{\includegraphics[width=2.96in]{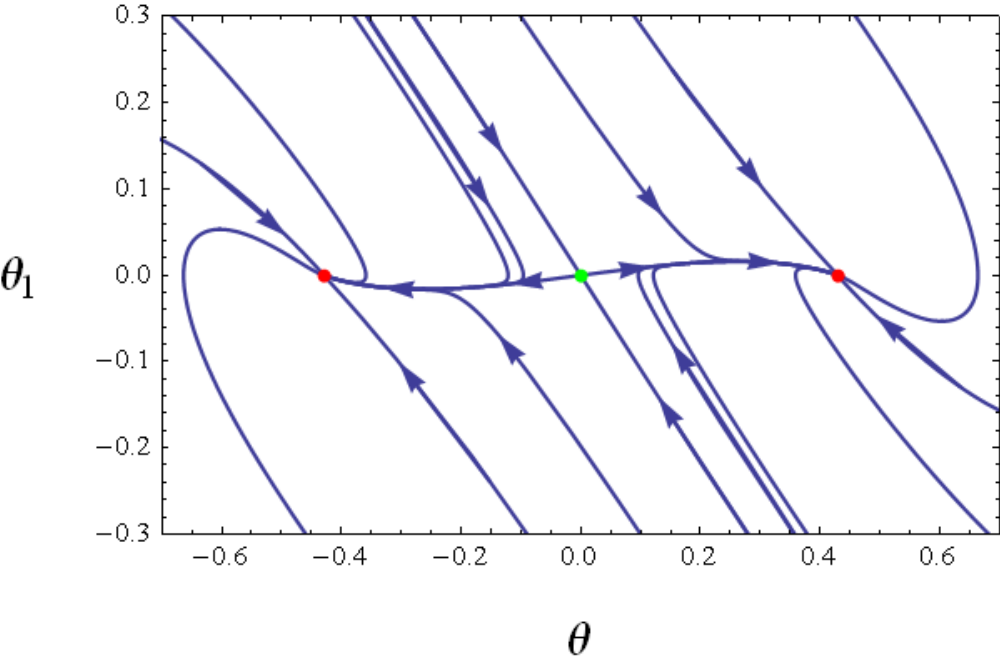}}
\hspace{1cm}
\subfigure[~stable node, $\mu = 2.5$]
{\includegraphics[width=2.8in]{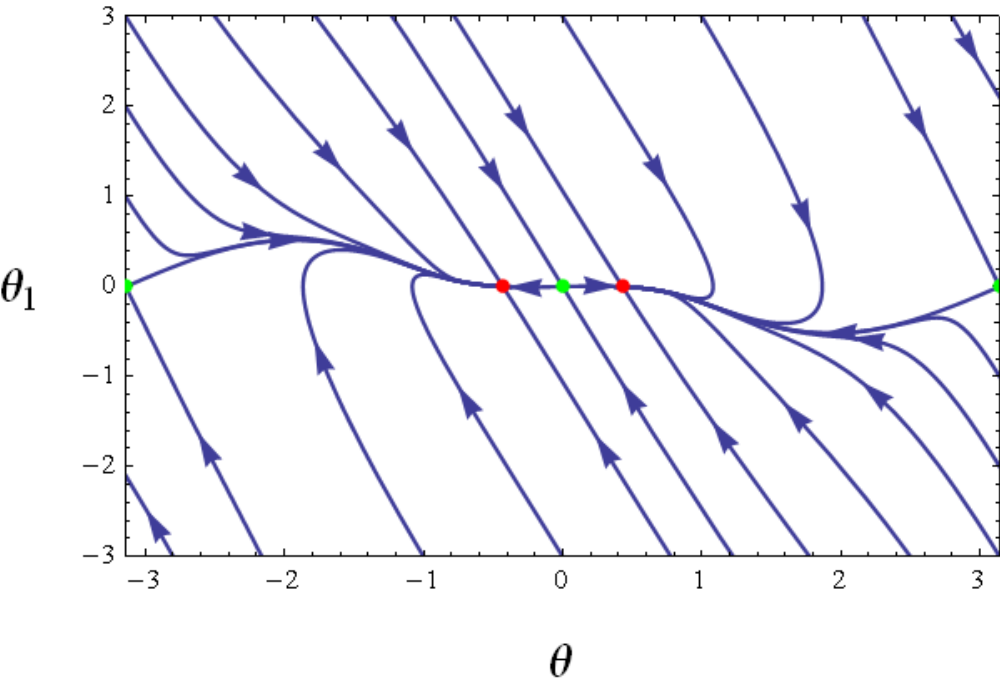}}
\caption{Phase trajectories around $(0,0)$ for $k=1.1$ and $\mu > 2\sqrt{k - \frac{1}{k}}$.}
\label{nodeomega1}
\end{figure}

For $0 \leq \mu<2\sqrt{k-\frac{1}{k}}$, 
$(\Omega_1,0)$ is a stable spiral with eigenvalues given by, 
\begin{equation}
\nonumber \lambda = -\frac{\mu}{2}\pm i\sqrt{(k-\frac{1}{k})-\frac{\mu^2}{4}}
\end{equation}
Trajectories spiral in with an angular frequency
$\nu \approx\sqrt{(k-1/k)-\mu^2/4}$, while their radial distance 
decreases as $e^{-\mu t/2}$. As $\mu\to 0$, this decay rate 
vanishes and ($\Omega_1,0)$ turns into a center (Figure \ref{spiralomega1}).
Also, $\nu$ vanishes as 
$\mu \to 2\sqrt{k-1/k}^-$, representing a smooth transition 
to a stable node, similar to the behaviour of the fixed point $(0,0)$.
When $\mu > 2\sqrt{k-\frac{1}{k}}$, 
$\Gamma^2 - 4\Delta>0$, hence $(\Omega_1,0)$ is a stable node 
(Figure \ref{nodeomega1}), 
with eigenvalues and eigenvectors given by,
\begin{equation}
\nonumber \lambda_{1,2} = \frac{-\mu \pm \xi_3}{2},\;\;\;
\mathbf{v}_{1,2} = \left(
\begin{array}{c}
1 \\
(-\mu \pm \xi_3)/2
\end{array} \right ) \; ,
\end{equation}
where $\xi_3 = \sqrt{\mu^2 - 4(k-1/k)}$.
Both $\lambda_1$ and $\lambda_2$
approach the value $-\sqrt{k- 1/k}$ as $\mu\to 2\sqrt{k- 1/k}^+$, indicating a 
stable degenerate node.
For $\mu=2\sqrt{k-\frac{1}{k}}$, $\Gamma^2 - 4\Delta=0$. 
In the linear stability analysis, $(\Omega_1,0)$ is a stable degenerate 
node with a single eigenvector,
\beq
{\mathbf{v}}= \left(
\begin{array}{c}
1 \\
-\sqrt{k-1/k} 
\end{array} \right) \; ,
\eeq
corresponding to the eigenvalue $\lambda=-\sqrt{k-1/k}$. However, 
degenerate nodes can be transformed 
into stable nodes or stable spirals due to perturbation introduced by
nonlinear terms. 

\subsection{Nature of $(\Omega_1,0)$ with nonlinearities}
As discussed in subsection \ref{nlin0}, (\ref{fo1}) and (\ref{fo2}) 
may be transformed 
to equations in $r$ and and $\phi$ with the substitutions,
$\theta-\Omega_1=r\cos \phi$ and $\theta_1=r\sin \phi$.
Define,
\begin{equation}
f(\phi) = \lim_{r \to 0}\phi^{'} = -\sqrt{k-\frac{1}{k}}\sin (2\phi) - \frac{1}{2}(k-\frac{1}{k} -1)\cos (2\phi) - \frac{1}{2}(k-\frac{1}{k} +1)
\end{equation}

$f(\phi)$ is negative at all points on the $\phi$ axis except at the 
fixed points given by $(0,\phi^*)$ with $\phi^*=n\pi -\alpha$, 
where $\alpha= \tan^{-1}\sqrt{k-1/k}$, ($n = 0, 1, 2, \dots $), where it is zero. These fixed points are separated by $n\pi$. Then, by the same reasoning as 
used in \ref{nlin0}, we can argue that $(0,0)$ cannot be a stable node. 
The remaining possibilities are a spiral or a degenerate node.

Let us consider the fixed point $(0,-\alpha)$, and let 
$\epsilon=\phi+\alpha$. Then we may expand $r^{'}$ and $\epsilon^{'}$ 
upto $O(r)$,
\beqar
r^{'} = -\frac{1}{2}\Big(k-\frac{1}{k} +1\Big)\big(\sin (2\alpha) - \sin (2\epsilon)\big) r + O(r^2) \hspace{3cm}
\label{Oreq} \\
\epsilon^{'} = -\Big(k-\frac{1}{k} +1\Big)\sin^2 \epsilon - \frac{3}{2}r\sqrt{1-\frac{1}{r^2}}\cos^3 \alpha  \hspace{3.8cm} \nonumber \\
-\frac{9}{4}\sqrt{1-\frac{1}{k^2}}\cos \alpha\; 
\sin (2\alpha) \;r\;\epsilon + O(r^2) \hspace{4.2cm}
\label{Oepeq}
\eeqar

We choose an initial point, $(r_0,\epsilon_0)$, and 
$0<r_0\ll \epsilon_0< min\{1,\alpha\}$. This choice ensures that both 
$r$ and $\epsilon$ will start decreasing. In the course of this monotonic decay, $r$ cannot reach zero before 
$\epsilon$ becomes zero, as there is a straight line trajectory moving 
downward along the $\phi$ axis. From (\ref{Oreq}) and (\ref{Oepeq}) we note that $r^{'}$ and $\epsilon^{'}$ 
each contains an $\epsilon$ independent term of $O(r)$. Therefore, as $r$ and 
$\epsilon$ decrease toward their respective zero values, the 1st order 
approximation gets even better. However, if at some stage, $\epsilon \sim r$, then the $O(r^2)$ terms would contribute on the same scale as some of the terms of $O(r \epsilon)$ in (\ref{Oreq}) and (\ref{Oepeq}). But the following argument rules out such a possibility.

\begin{figure}[!h]
\centering
\subfigure[~stable degenerate node]
{\includegraphics[width=2.8in]{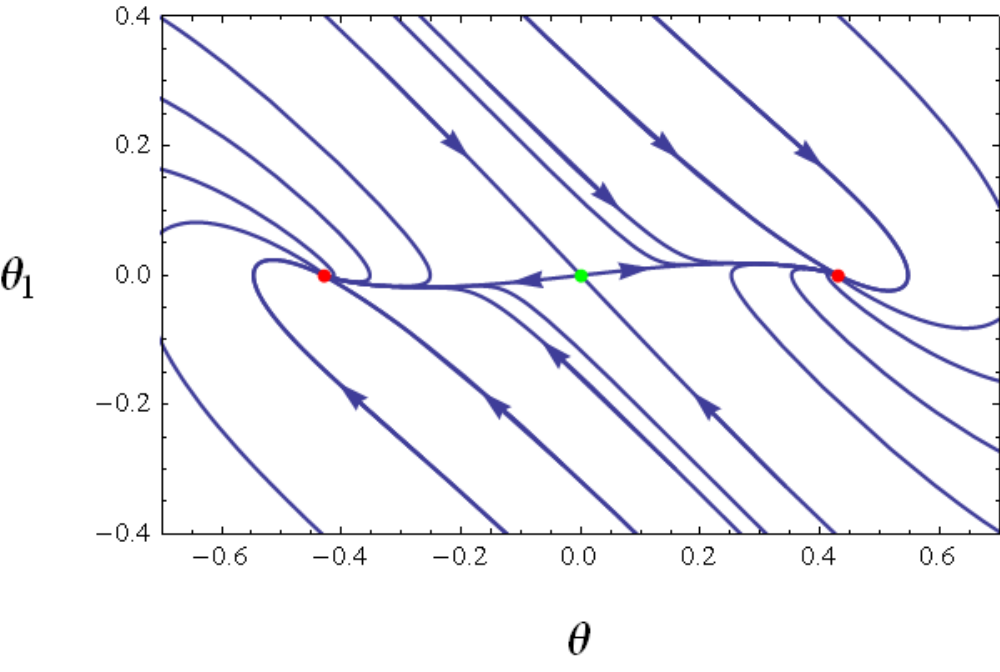}}
\hspace{1cm}
\subfigure[~region around $(\Omega_1,0)$ magnified]
{\includegraphics[width=2.8in]{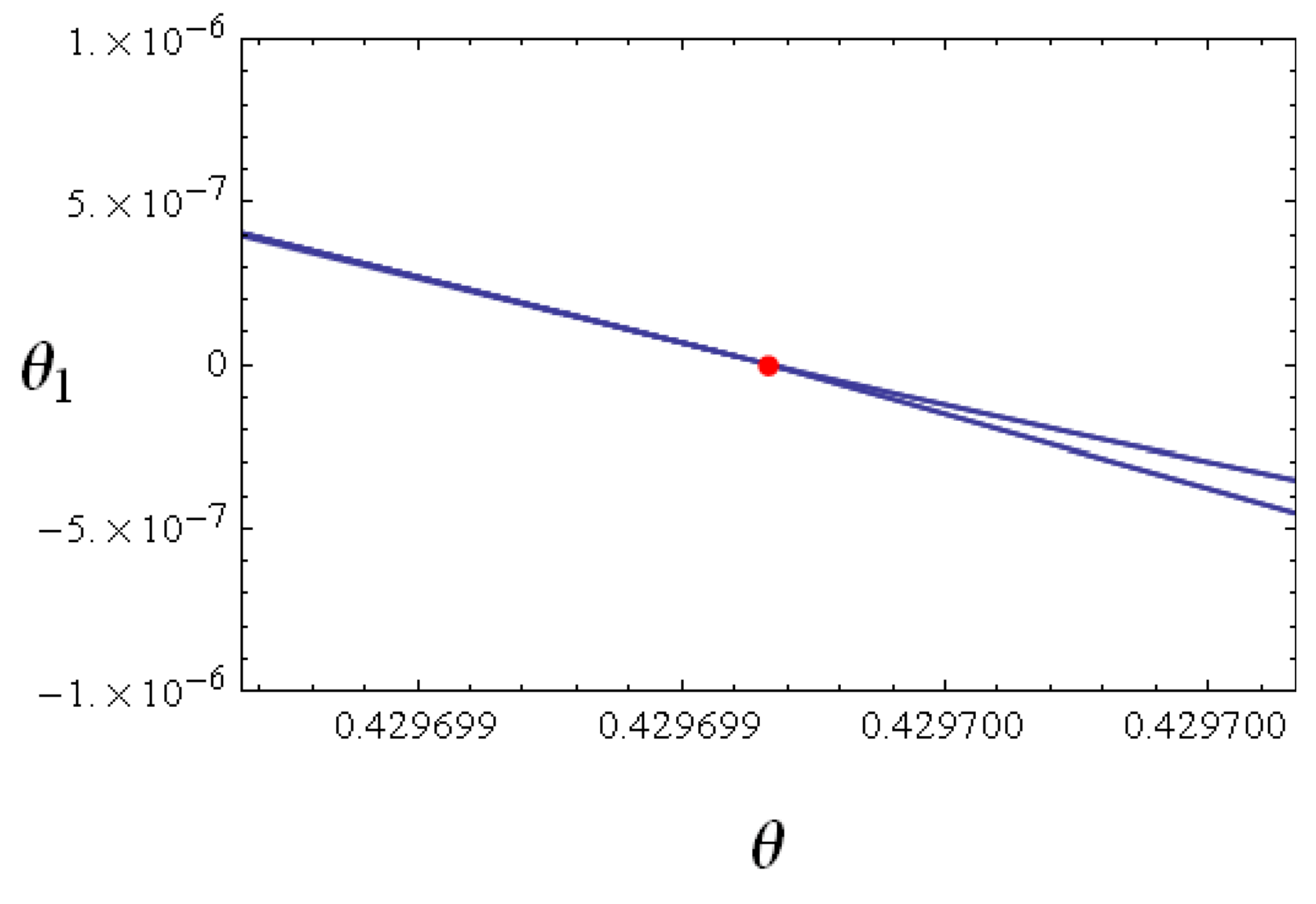}}
\caption{Phase trajectories for $k=1.1$ and $\mu = 2\sqrt{k - \frac{1}{k}}$.}
\label{degomega1}
\end{figure}

Let us consider a specific case and choose $r_0=\epsilon_0^{20}$. 
If $\epsilon$ is to become $O(r)$, at some stage, 
we must have $r=\epsilon^2$. However, for $r=\epsilon^2$, we have
\begin{eqnarray}
\nonumber r^{'} &=& -\sqrt{k-\frac{1}{k}} \epsilon^2 + O(\epsilon^3) \\
\nonumber \epsilon^{'} &=& -\left[(k-\frac{1}{k} +1)+\frac{3}{2}
\frac{\sqrt{1-\frac{1}{k^2}}}{(k-\frac{1}{k} +1)^\frac{3}{2}}\right] 
\epsilon^2 + O(\epsilon^3)
\end{eqnarray}
whereas along the curve $r=\epsilon^2$, $dr/d\epsilon=2\epsilon$. 
Therefore, for a given value of $k>1$, we can always select an 
$\epsilon_0$, sufficiently small, for which, at all points on the 
curve $r=\epsilon^2$ contained between $\epsilon=0$ and $\epsilon=\epsilon_0$,
the ratio $r^{'}/\epsilon^{'} > dr/d\epsilon$. This would guarantee that 
the trajectory cannot penetrate down this curve, which means 
that $\epsilon$ cannot reach zero before $r$ does. Thus, for a suitable choice of initial conditions, the trajectory must 
slow to a halt at $(r=0,\epsilon=0)$. In the $\theta-\theta_1$ plane, 
this means that there exist trajectories 
which start at a finite distance from $(\Omega_1,0)$ and reach it 
along the line of slope $-\alpha$. Also, there is no other such line with 
a different slope. Hence, $(\Omega_1,0)$ is a stable degenerate node (Figure \ref{degomega1}).
Figure \ref{param}b gives the nature of the fixed point at 
$(\pm\Omega_1,0)$ in different parts of the parameter space.

\section{Nature of the fixed point $(\pi,0)$}
\label{naturepi}
The Jacobian matrix at ($\pi$,0) is given as,
\beq
\mathbf{J}(\pi,0) = \left(
\begin{array}{ccc}
0 & 1 &  \\
k+1 & -\mu  
\end{array} \right) \; .
\eeq
The fixed point is a saddle for all values of $k$ and remains so 
even with the inclusion of nonlinear terms.
The eigenvalues and corresponding eigenvectors are given by,
\beq
\lambda_{1,2} = \frac{-\mu \pm \xi_2}{2},\;\;\;
\mathbf{v}_{1,2} = \left(
\begin{array}{c}
1 \\
(-\mu \pm \xi_2)/2
\end{array} \right ) \; , \nonumber \\
\eeq
where $\xi_2 = \sqrt{\mu^2 + 4(k+1)}$.
\subsection{Trajectories}
Damping of the bead leads to some qualitatively different trajectories 
in addition to those observed for the frictionless case\cite{shovan1}. 
These are mainly 
the different kinds of damped oscillations (underdamped, critically damped, 
overdamped) about the stable equilibrium points.
Some of these are illustrated with the following numerical plots.
\begin{figure}[!h]
\centering
\subfigure[~$k=0.75$, $\mu=0.05$, $\theta^{'}(0)=0$]
{\includegraphics[width=2.0in]{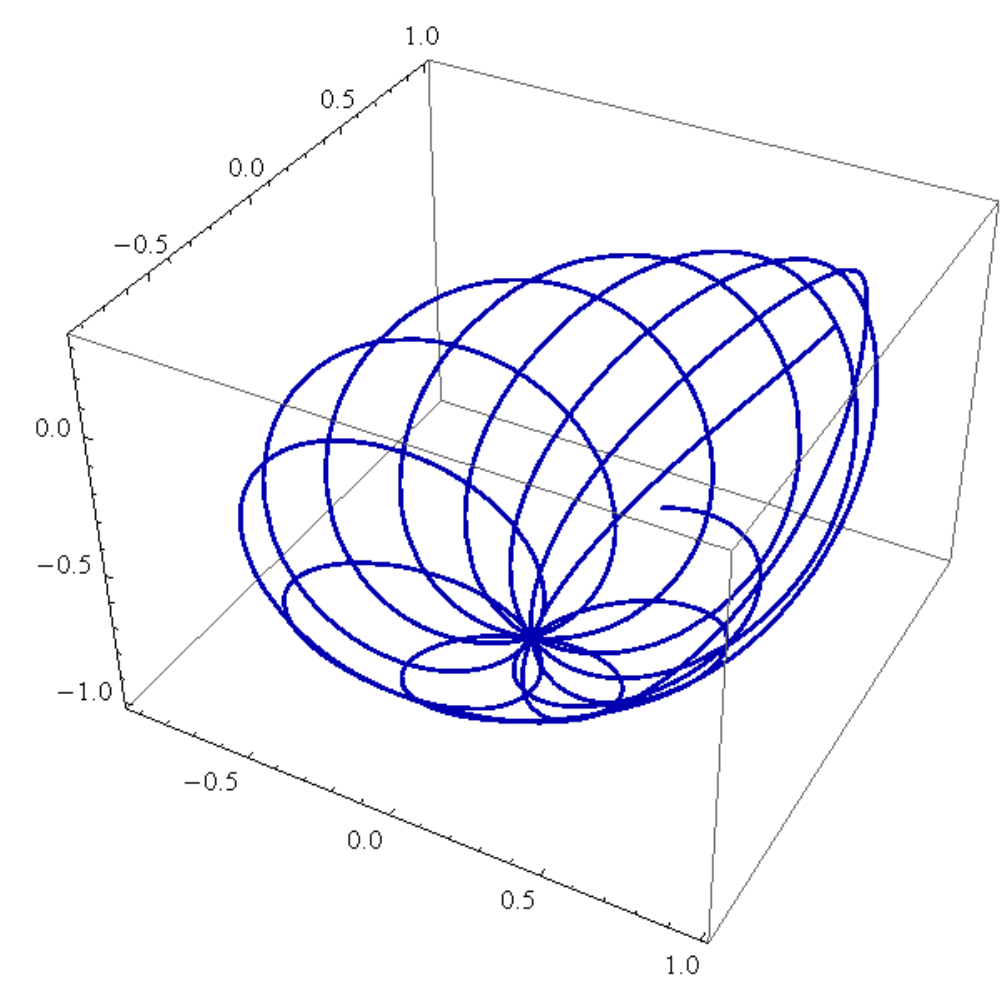}}
\hspace{0.7cm}
\subfigure[~$k=4$, $\mu=0.5$, $\theta^{'}(0)=0$]
{\includegraphics[width=2.0in]{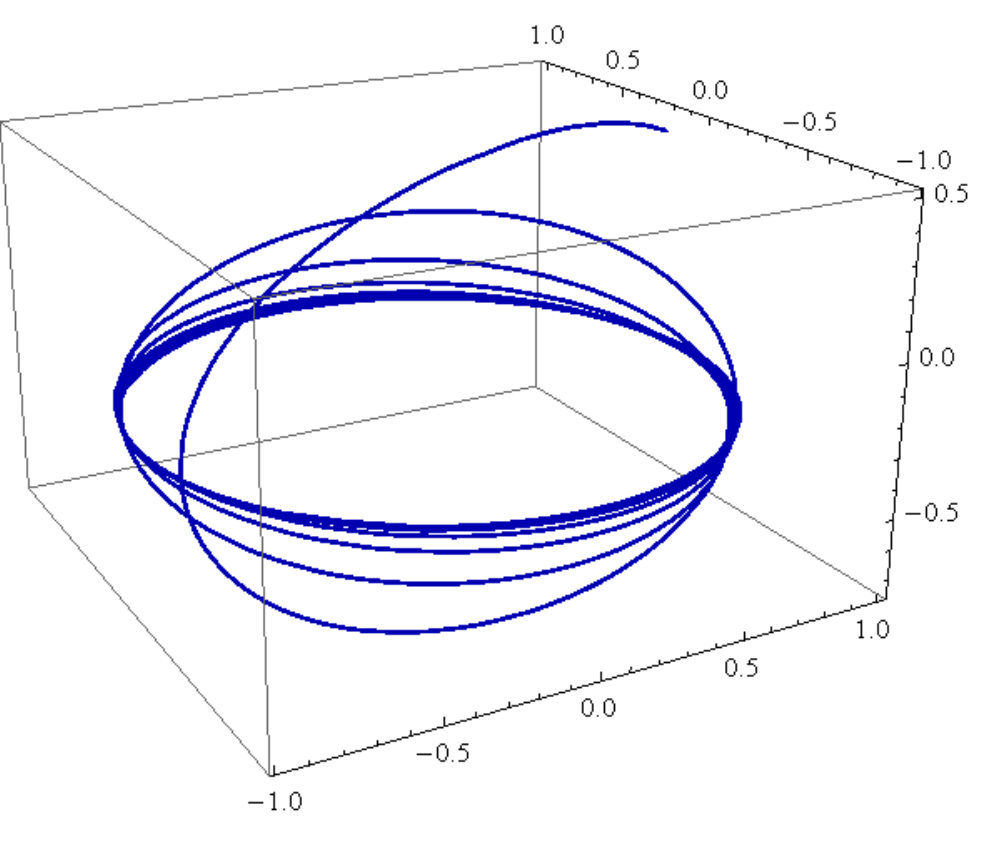}}
\caption{Underdamped oscillation about (a) $\theta=0$ and (b) 
$\theta = \Omega_1$.}
\label{traj1}
\end{figure}
\begin{figure}[!h]
\centering
\subfigure[~$k=0.75$, $\mu=1.5$, $\theta^{'}(0)=0$]
{\includegraphics[width=2.0in]{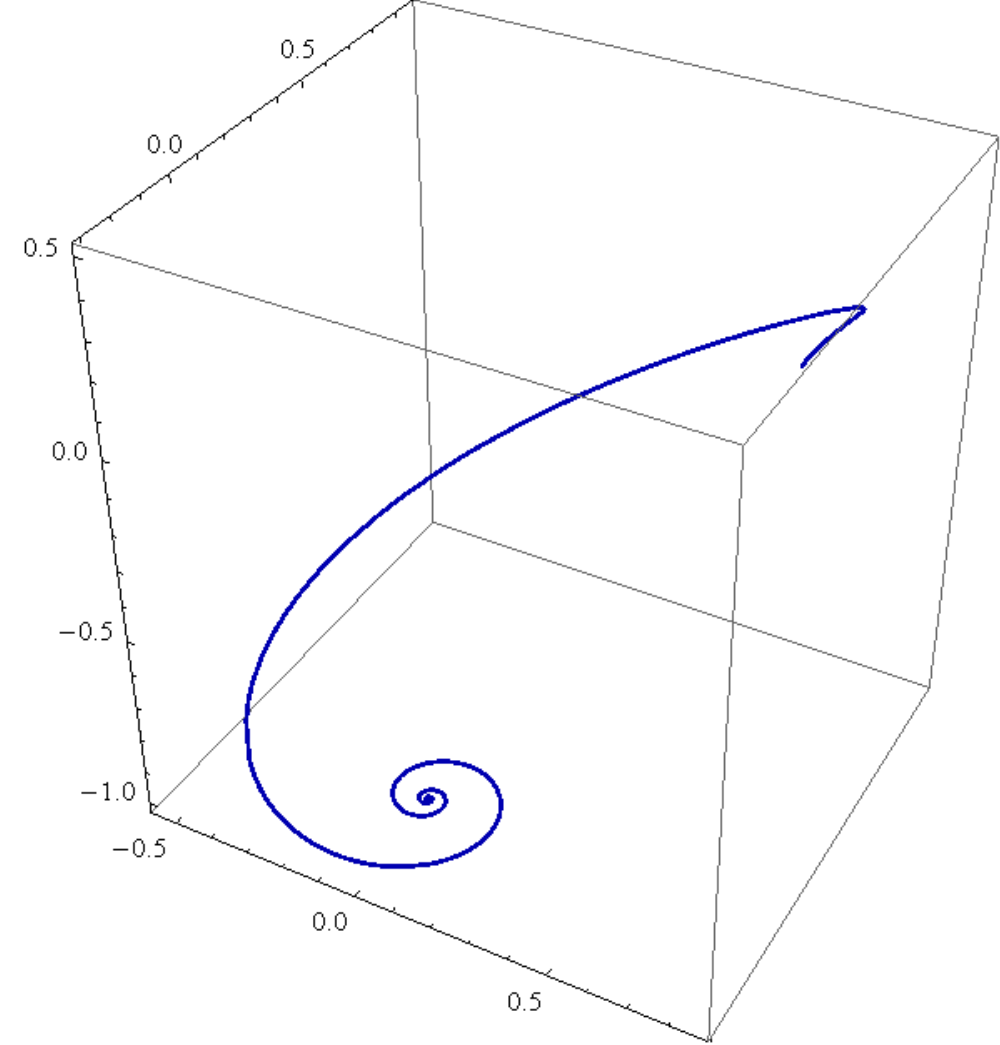}}
\hspace{1cm}
\subfigure[~$k=4$, $\mu=4$, $\theta^{'}(0)=0$]
{\includegraphics[width=2.0in]{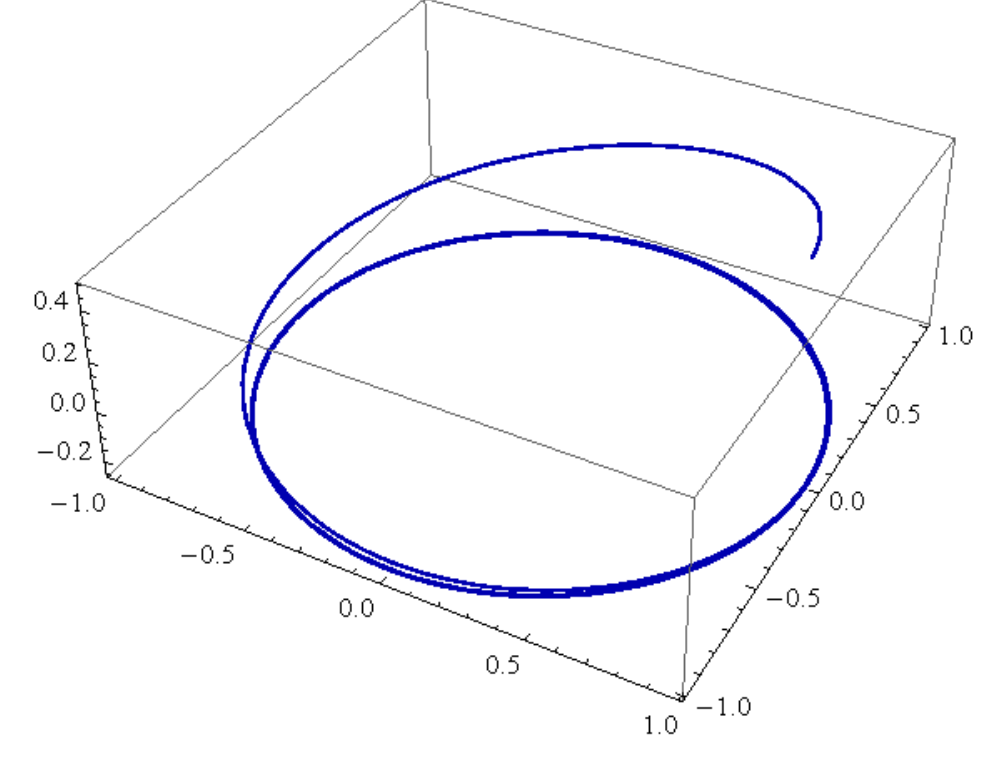}}
\caption{Overdamped oscillation about (a) $\theta=0$ (b) $\theta = \Omega_1$.}
\label{traj2}
\end{figure}
\section{Phase portraits and Bifurcation}
For $0 \le k < 1$, the fixed point at $(0,0)$ transforms its nature 
as the damping coefficient is varied.
It is a center at $\mu=0$, as $\mu$ increases, it becomes a stable spiral.
At $\mu=2\sqrt{1-k}$, it turns into a stable degenerate node. It makes 
a smooth transition to a stable node as damping is increased further. 
Thus, a spiral-node bifurcation takes place at this critical condition 
(Figures \ref{sspiral0} and \ref{deg}). 
Physically, 
as damping is gradually increased from $0$, the system undergoes a continuous 
transition from undamped oscillations of the bead about $\theta=0$ (center),
to underdamped oscillations
(stable spiral). At $\mu=2\sqrt{1-k}$, the system is critically damped
(degenerate node) and becomes overdamped (stable node) as $\mu$ is increased
further. 
\begin{figure}[!h]
\centering
\subfigure[~unstable spiral, $\mu=-1$]
{\includegraphics[width=1.9in]{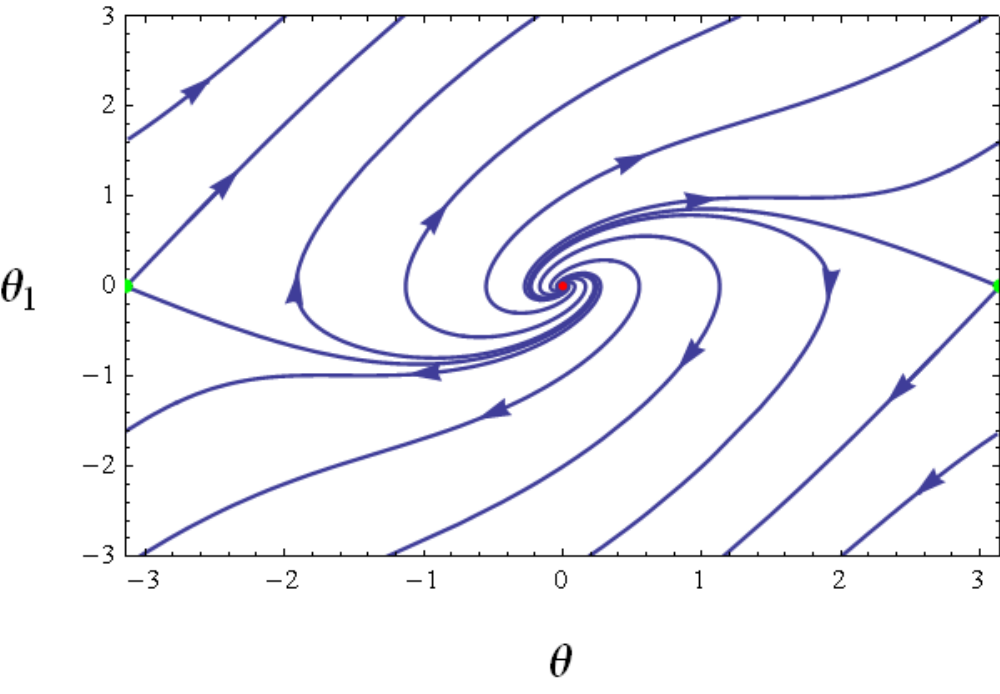}}
\hspace{0.7cm}
\subfigure[~center, $\mu=0$]
{\includegraphics[width=1.8in]{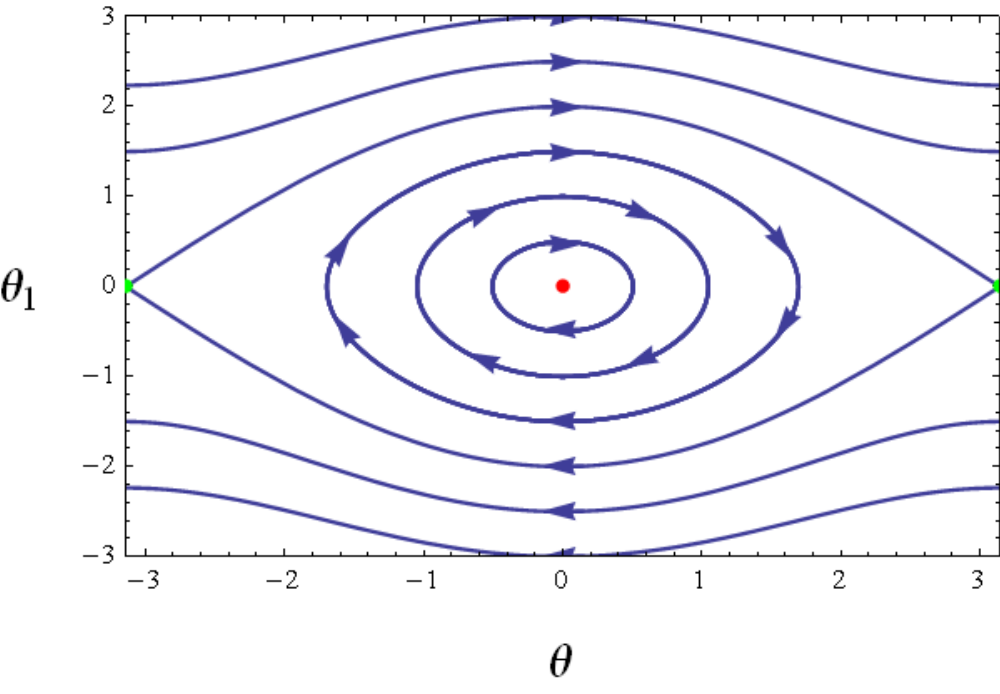}}
\hspace{0.7cm}
\subfigure[~stable spiral, $\mu = 1$]
{\includegraphics[width=1.9in]{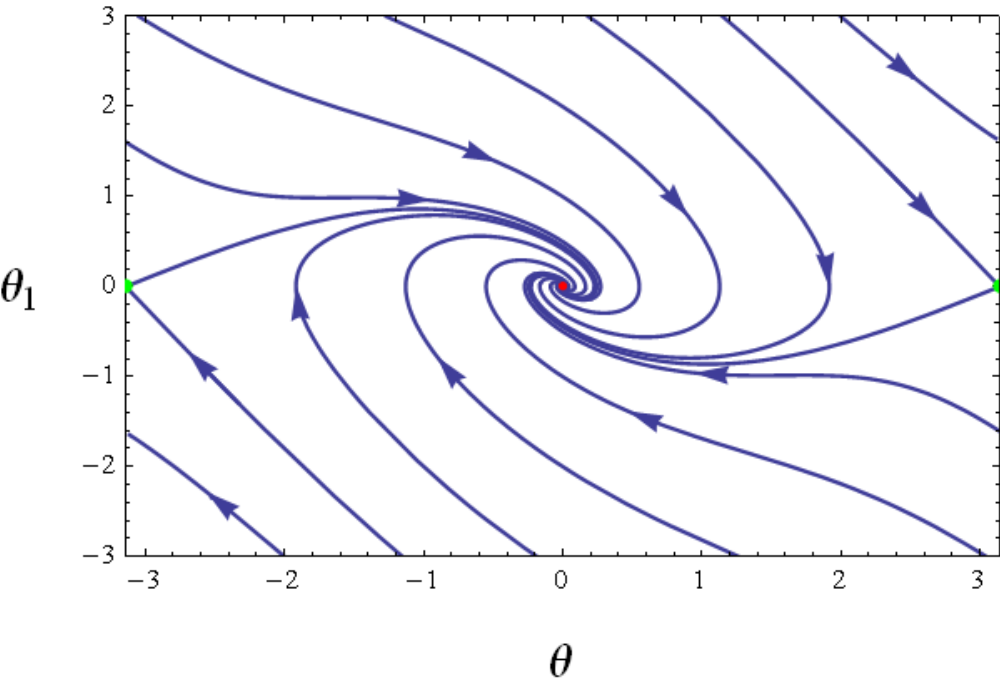}}
\caption{Phase portraits for $k=0$ showing degenerate Hopf bifurcation.}
\label{Hopf1}
\end{figure}

For negative damping, $(0,0)$ becomes an unstable spiral and changes to
an unstable node as $\mu$ is made more negative.
Consequently, as one crosses $\mu=0$, the fixed point $(0,0)$, undergoes a
degenerate Hopf bifurcation (Figure \ref{Hopf1}).

\begin{figure}[h]
\centering
\subfigure[~unstable node, $\mu=-0.2$]
{\includegraphics[width=2.8in]{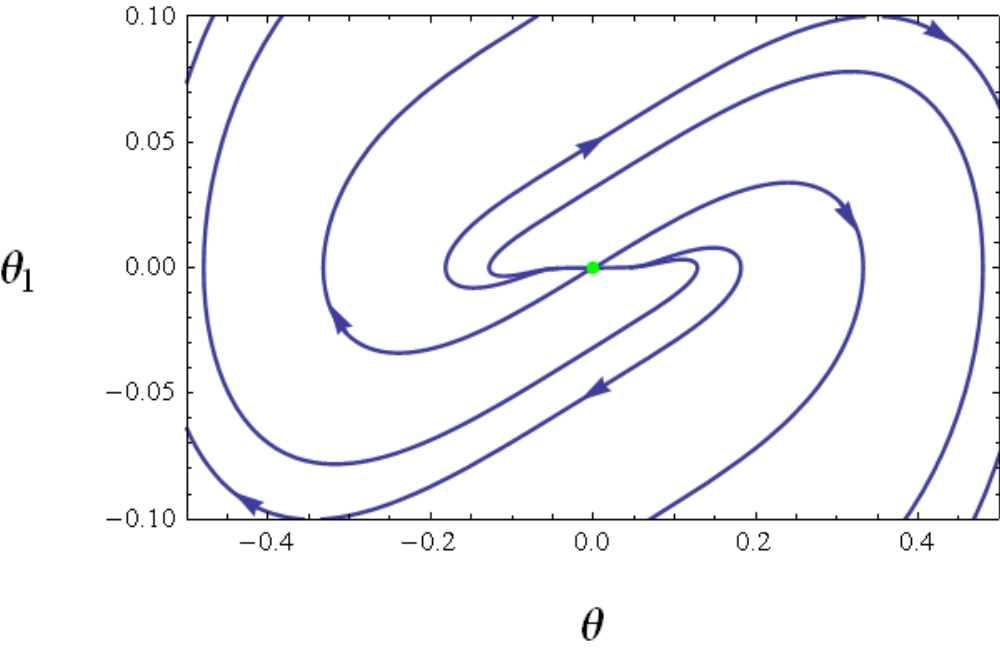}}
\hspace{0.7cm}
\subfigure[~stable node, $\mu=0.2$]
{\includegraphics[width=2.8in]{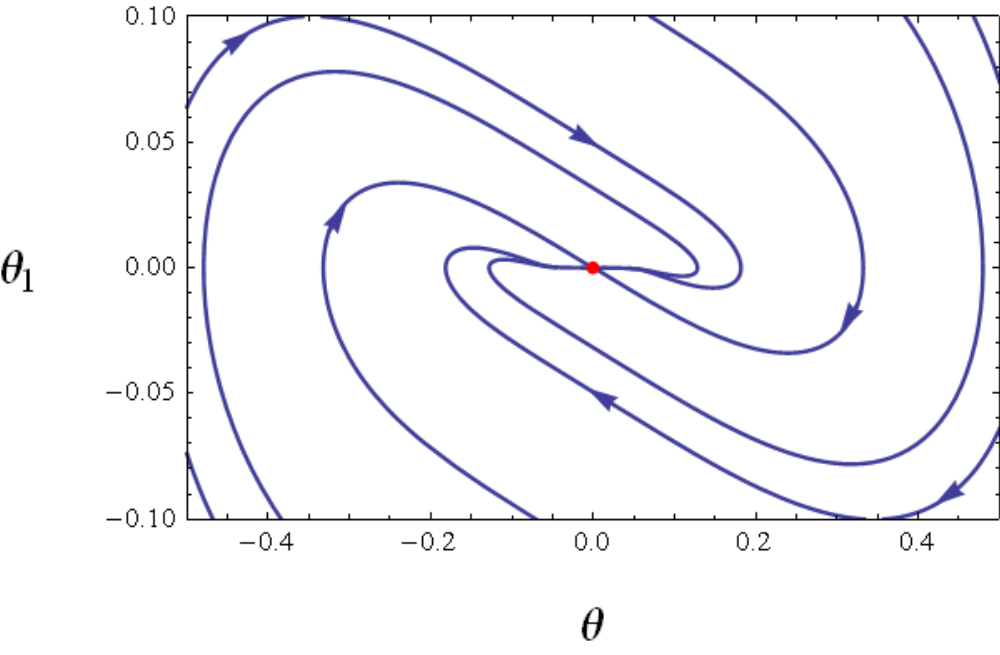}}
\caption{Phase portraits for $k=1$.}
\label{k_1}
\end{figure}

With increase in the angular speed of the hoop (i.e., $k$), the stability of 
the origin degrades continuously. When $k=1$, $(0,0)$ is a weak center.
A special case of Hopf bifurcation occurs, when $\mu$ is swept from negative to
positive values acroos 0, keeping $k$ fixed at 1 (Figure \ref{k_1}).

\begin{figure}[h]
\centering
\subfigure[~Supercritical bifurcation at $k=1$, $\mu=0$]
{\includegraphics[width=2.8in]{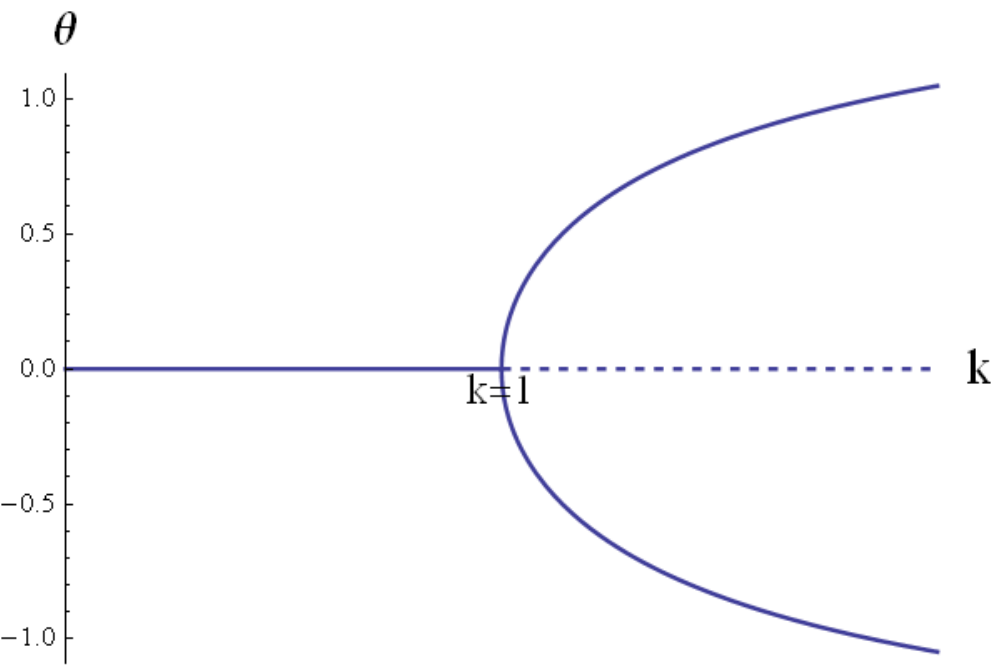}}
\hspace{1cm}
\subfigure[~Supercritical bifurcation at $k=1$, $\mu=\sqrt{2}$]
{\includegraphics[width=2.8in]{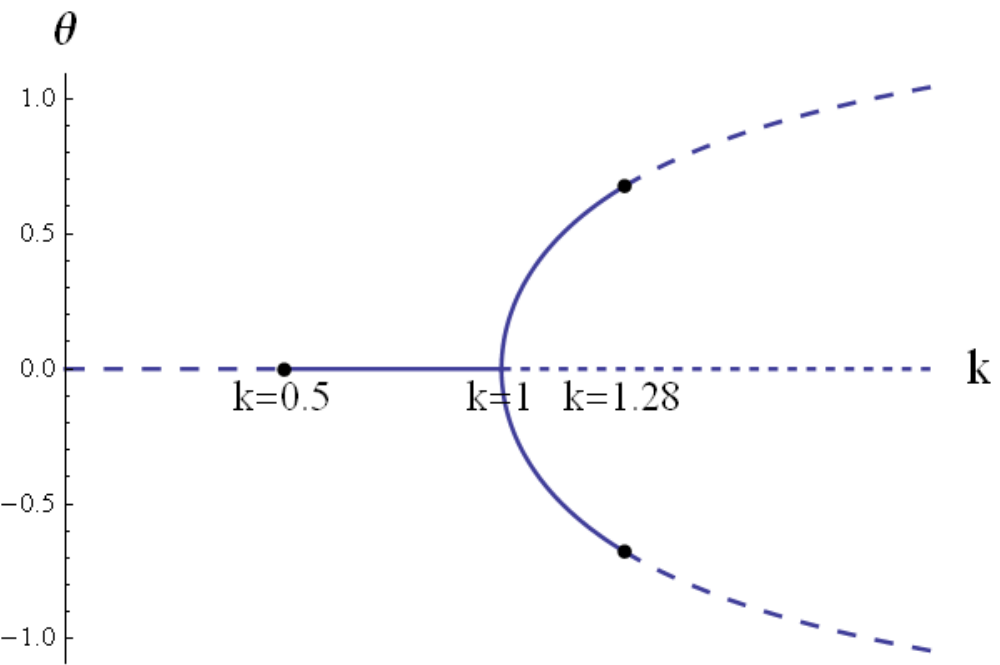}}
\caption{Section of the bifurcation diagram for (a) $\mu=0$ and (b) $\mu >0$.
In (a) solid curve represents center, dashed curve represents saddle. 
In (b) solid curve represents stable node, densely dashed curve represents 
saddle and sparsely dashed curve represents stable spiral. 
Supercritical pitchfork bifurcation occurs at $k=1$ and spiral-node 
bifurcation occurs at $k=0.5$ and $k=1.28$.}
\label{pitchfork}
\end{figure}

As $k$ is increased beyond $1$,
$(0,0)$ transforms from a stable ($\mu > 0$) or unstable ($\mu <0$) node
to a saddle. 
Two new stable nodes appear at $\pm \Omega_1 = \pm \cos^{-1}(1/k)$
and branch out in opposite directions.
Thus, a supercritical pitchfork bifurcation occurs 
at \{$k=1$\} (Figure \ref{pitchfork}a). 

As $k$ increases from $1$ to $\infty$, $\Omega_1$ varies from $0$ to $\pi/2$. 
The fixed point $(\Omega_1,0)$, is a center 
for zero damping, a stable spiral in the region $0<\mu<2\sqrt{k-1/k}$ 
(underdamped oscillation), and becomes a stable degenerate node
at critical damping $\mu=2\sqrt{k-1/k}$. For the overdamped condition 
$\mu>2\sqrt{k-1/k}$, it is a stable node.

\begin{figure}[h]
\centering
\subfigure[~unstable spiral, $\mu = -1$]
{\includegraphics[width=1.9in]{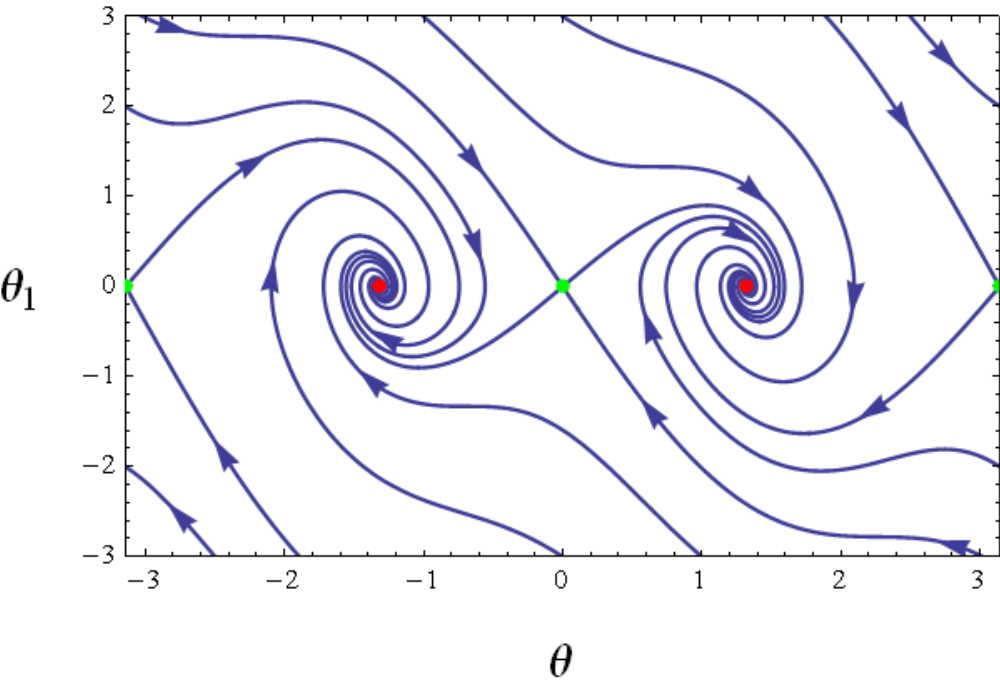}}
\hspace{0.7cm}
\subfigure[~center, $\mu = 0$]
{\includegraphics[width=1.8in]{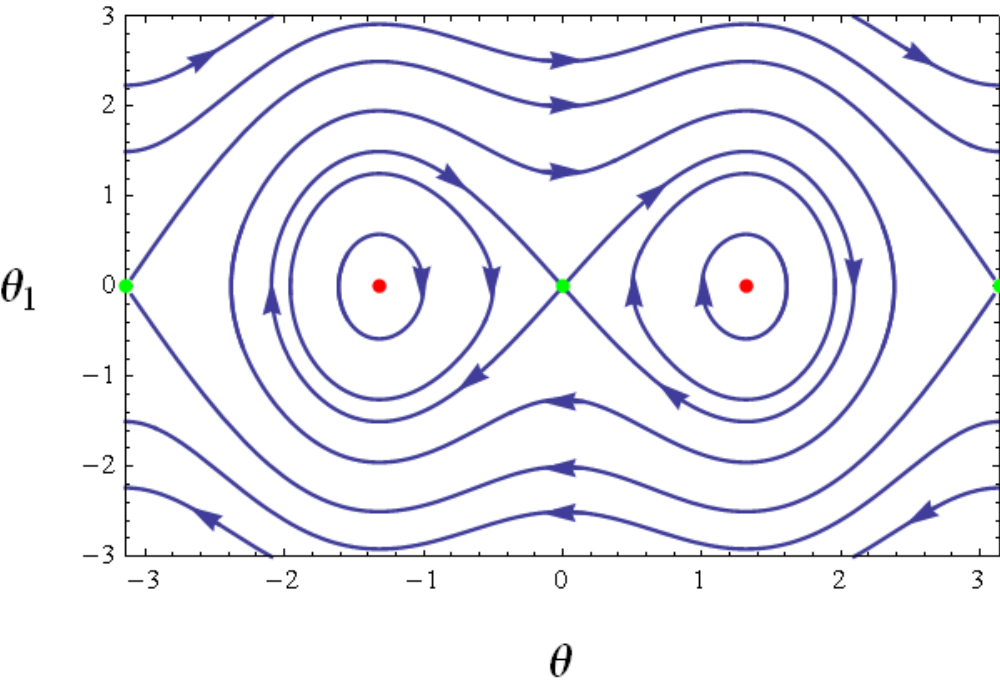}}
\hspace{0.7cm}
\subfigure[~stable spiral, $\mu = 1$]
{\includegraphics[width=1.9in]{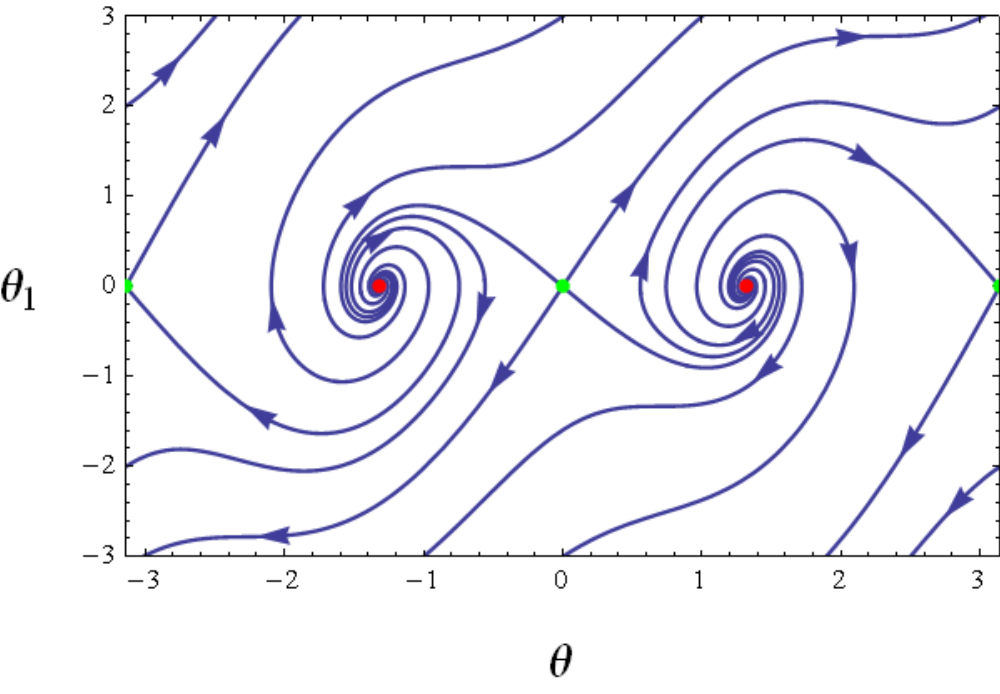}}
\caption{Phase portraits for $k=4$ showing degenerate Hopf bifurcation.}
\label{Hopf2}
\end{figure}

For negative damping, we get just the unstable counterparts. 
Accordingly, a spiral-node bifurcation is observed at 
$\mu=\pm 2\sqrt{k-1/k}$ (Figures \ref{spiralomega1}, 
\ref{nodeomega1}, \ref{degomega1} and \ref{pitchfork})b. A degenerate 
Hopf bifurcation is observed for $k > 1$ and $\mu =0$ 
(Figure \ref{Hopf2}).
The fixed points $(\pm\pi,0)$ are saddles for all values of $k$.
They have saddle connections between them at
$\mu =0$, which break in opposite directions for positive and 
negative damping. 

The above observations are summarized in Figure \ref{bif_dia} and
Table \ref{BT} below.
\begin{figure}[h]
\centering
{\includegraphics[width=8cm]{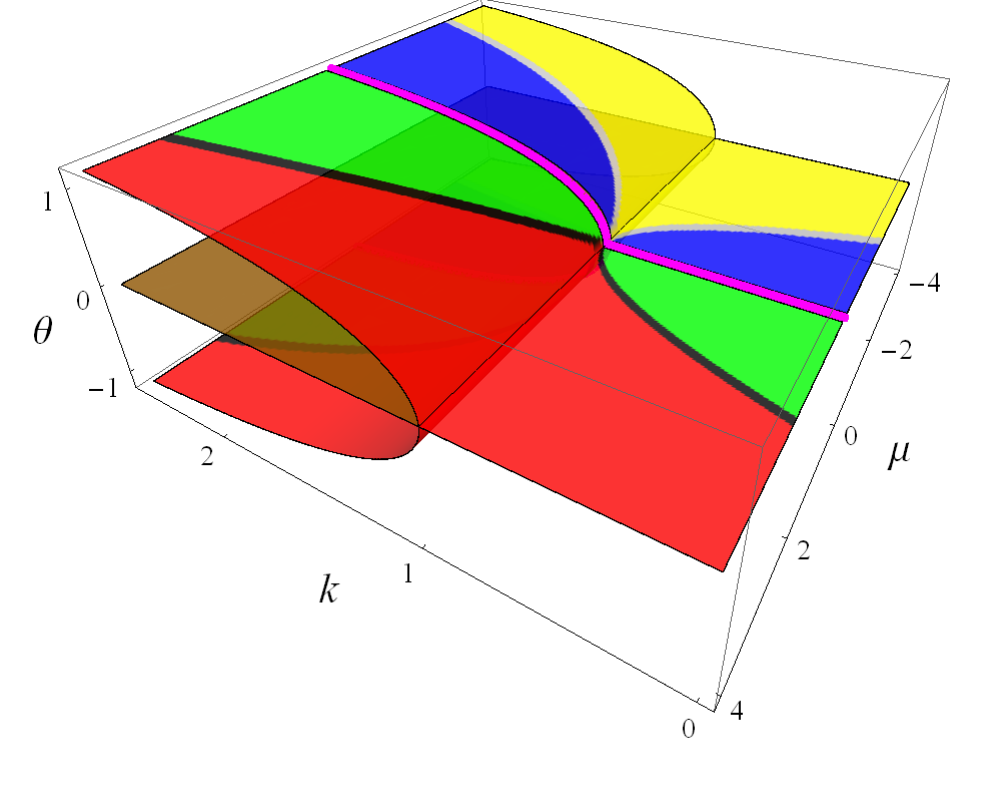}}
\caption{The bifurcation diagram}
\label{bif_dia}
\end{figure}

In Fig \ref{bif_dia}, red denotes stable node, green denotes 
stable spiral, blue denotes unstable spiral, yellow denotes unstable node, 
brown denotes saddle, pink denotes center, gray denotes unstable degenerate 
node, and black denotes stable degenerate node.

\begin{table}[ht]
\centering
\caption{\label{BT} Bifurcation Table.}
\vspace{0.2cm}
{\renewcommand{\arraystretch}{1.5}
\renewcommand{\tabcolsep}{0.2cm}
\begin{tabular}[c]{|l|c|c|c|}
\hline
Points in parameter space & Bifurcation along $k$ & Bifurcation along $\mu$ & Figure references\\
\hline
1) $(k,0)$ ; $k\neq 1$ & -- & degenerate Hopf & Figures \ref{Hopf1}, \ref{Hopf2} \\
\hline
2) $(k,\pm2\sqrt{1-k})$ ; $0\leq k<1$ & spiral-node & spiral-node & 
Figures \ref{sspiral0}-\ref{deg}, \ref{pitchfork}b \\
\hline
3) $(1,0)$ & supercritical pitchfork & Hopf & Figures \ref{Hopf1}b, 
\ref{spiralomega1}a, \ref{k_1} , \ref{pitchfork}a\\
\hline
4) $(1,\mu)$ ; $\mu \neq0$ & supercritical pitchfork & -- & Figures \ref{sspiral0}b, \ref{nodeomega1}a, \ref{k_1}b, \ref{pitchfork}b\\
\hline
5) $(k,\pm2\sqrt{k-\frac{1}{k}})$ ; $k>1$ & spiral-node & spiral-node & Figures \ref{spiralomega1}-\ref{degomega1}, \ref{pitchfork}b \\
\hline
\end{tabular}}
\end{table}

Up to now, we have limited attention to those bifurcations resulting from a variation of k or a variation of $\mu$. From Figure \ref{bif_dia}, we see that the curves $\mu^2 = 2(1 - k)$, $\mu^2=2(k - 1/k)$, and $k=1$ divide the parameter space into $8$ distinct regions of different dynamics. All these regions meet at the point $\{k=1,\mu=0\}$. Traversing suitable curves in $k-\mu$ space, one can move from any one region to another, yielding new kinds of bifurcation. Following such a curve amounts to keeping a certain function $\alpha (k, \mu)$ constant, while varying some other function $\beta (k, \mu)$. Mathematically, the possibilities are rich. But whether it is possible to actually implement this in the bead-hoop system is subject to further inquiry. However, this would attain physical significance if there exists another system where $\alpha$ and $\beta$ themselves are the control parameters.

\section*{Concluding Remarks}
The simple introduction of damping to the bead-hoop system enriches its dynamics and leads to various new modes of motion and different classes of bifurcations. We have studied this system over the entire parameter space and presented phase portraits and trajectories.
This serves to illustrate the qualitative changes in the system's dynamics across different bifurcation curves. We have presented exact analytical treatment of the borderline cases where linearization fails, for which no general methods are available in the literature. The method of transforming to polar coordinates and using order of magnitude arguments, employed in this article, can serve as a useful technique for other dynamical systems as well.

\end{document}